\documentclass[useAMS,usenatbib]{mn2e}
\usepackage{graphicx}
\usepackage{amsmath,amssymb}             
\usepackage{amssymb}
\usepackage{amsmath}
\usepackage{amsfonts}
\usepackage{soul}
\usepackage{color}

\def\AA{Astron. \& Astroph.}

\def\ARAA{Ann. Rev. Astron. Astroph.}

\def\ApJ{Astrophys. J.}

\def\ApJS{Astrophys. J. Suppl.}
\def\AJ{Astron. J.}
\def\NewA{New Astron.}
\def\Nat{Nature}

\def\PASJ{Publ. Astron. Soc. Japan}
\def\PhRL{Phys. Rev. Lett.}

\def\MN{Mon. Not. Royal Astron. Soc.}

\title[Evolution of growing black holes in axisymmetric galaxy cores]{Evolution of growing black holes in axisymmetric galaxy cores}
\author[J. Fiestas, O. Porth, P. Berczik, R. Spurzem] {
J. Fiestas$^{1,3}$ \thanks{E-mail:fiestas@ari.uni-heidelberg.de}, 
O. Porth $^{2}$, 
P. Berczik $^{1,3,5}$ and 
R. Spurzem$^{3,1,4}$ 
	\footnotemark[1]{}\\
$^{1}$Astronomisches Rechen-Institut, M\"{o}nchhofstra\ss e 12-14,   D-69120 Heidelberg, Germany\\
$^{2}$Max-Planck-Institut f\"ur Astronomie K\"{o}nigstuhl 17 D-69117 Heidelberg\\
$^{3}$National Astronomical Observatories of China, Chinese Academy of Sciences NAOC/CAS,\\ \ 20A Datun Rd., Chaoyang District, Beijing 100012, China\\
$^{4}$ The Kavli Institute for Astronomy and Astrophysics at Peking University\\
$^{5}$ Main Astronomical Observatory, National Academy of Sciences of Ukraine, MAO/NASU,\\ \ \ 27 Akademika Zabolotnoho St. 03680 Kyiv, Ukraine
}

\begin{document}

\date{}

\pagerange{\pageref{firstpage}--\pageref{lastpage}} \pubyear{2011}

\maketitle

\label{firstpage}

\begin{abstract}
NBody realizations of axisymmetric collisional galaxy cores (e.g. M32, M33, NGC205, Milky Way) with embedded growing black holes are presented. Stars which approach the disruption sphere are disrupted and accreted to the black hole. We measure the zone of influence of the black hole and disruption rates in relaxation time scales. We show that secular gravitational instabilities dominate the initial core dynamics, while the black hole is small and growing due to consumption of stars. Later, the black hole potential dominates the core, and loss cone theory can be applied. Our simulations show that central rotation in galaxies can not be neglected for relaxed systems, and compare and discuss our results with the standard theory of spherically symmetric systems.
\end{abstract}


\section{Introduction}

Galaxy cores are the hosts of supermassive black holes (SMBHs), the engines of quasars and of active galactic nuclei. There is increasing evidence that SMBHs play an important role in the formation and global evolution of galaxies. They are commonly observed at the centers of many nearby galaxies \citep{shankar09}, and the existence of quasars at least up to redshifts $z=6$ \citep{degraf10,willot10} implies that many of these SMBHs reached nearly their current masses at very early times. Evolution of galactic nuclei during and after the era of peak quasar activity therefore took place with the SMBHs already in place. The energy released by SMBHs during and after the quasar epoch must have had a major impact on how gas cooled to form galaxies and galaxy clusters \citep{scannapieco05}. However, the detailed history of SMBH growth is still being debated. Some work has focused on the possibility that the seeds of SMBHs were black holes of much smaller mass---either remnants of the first generation of stars, so-called ``Population III black holes'' \citep{MR-01}, or the (still speculative) ``intermediate-mass black holes'' (IMBHs), remnants of massive stars that form in dense clusters via physical collisions between stars \citep{portegies04,mapelli10}.

Observations with the Hubble Space Telescope have elucidated the run of stellar density near the centers of nearby galaxies \citep{ferrarese06,cote07,glass11}. Nevertheless, in the majority of galaxies massive enough to contain SMBHs, the central relaxation time is much greater than the age of the universe, due both to the (relatively) low stellar densities and also to the presence of a SMBH, which increases $v_{\rm rms}$ \citep{faber97,ferrarese06}. This long relaxation times imply that nuclear structure will still reflect the details of the nuclear formation process. Beyond the Local Group, essentially all of the galaxies for which the SMBH's influence radius is spatially resolved have 'collisionless' (non-relaxed) nuclei with low nuclear densities. The nuclei of these 'core' galaxies may have been much denser before the cores were created by probably binary SMBHs.

Only the smallest galaxies known to harbor SMBHs have nuclear relaxation times of $\sim 10$ Gyr or shorter. In such a nucleus the stellar distribution will have had time to evolve to a collisionally relaxed system. Galactic spheroids fainter than $M_V$ = -18 show this property. The Milky Way nucleus is 'collisional'. It has a half-mass relaxation time $t_{\rm rh} \sim 5 \times 10^{10}$ yr but the steep density profile implies $t_{\rm rh} \sim 6 \times 10^9$ yr at 0.2 $r_{\rm hm}$ (0.6 pc), where $r_{\rm hm}$ is the half-mass radius, and $\sim 3.5 \times 10^9$ yr at 0.1 $r_{\rm hm}$ (0.3 pc), assuming Solar-mass stars \citep{RPP06}. Collisional nuclei are present in three other Local Group galaxies (M32, M33 and NGC205) \citep{lauer98,valluri05} although M32 is the only one of these to exhibit dynamical evidence for a SMBH \citep{valluri05}.

Since the 1980s, the dominant model for the formation of elliptical galaxies and bulges -- the stellar systems that contain SMBHs -- has been the merger model \citep{toomre77}. An almost certain consequence of a merger is the in-fall of the progenitor galaxies' SMBHs into the nucleus of the merged system, resulting in the formation of a binary SMBH \citep{begelman80}, as observations of uncoalesced dual SMBHs show \citep{rodriguez06,valtonen08}. There is no generally agreed idea on whether and how fast a binary SMBH coalesces after a galaxy merger. Earlier work (mostly numerical simulations) in spherically symmetric nuclei discussed a stalling problem which would hang up the binary SMBH at some sub-pc separation (stalling or last parsec problem, cf. e.g. \cite{gould00, milos03, hemsendorf02, makino04, berczik05}). But it turns out that any degree of more realism helps clearing the last parsec problem, such as axisymmetry of the galaxy merger remnant \citep{berczik06,perets08,berentzen09,berczik11} or the presence of gaseous material in the nucleus \citep{dotti09,mayer10, callegari09, callegari10}. A self-consistent study of the combined effect of high-accuracy high-resolution stellar dynamics with binary black holes, together with evolution of a central galactic nuclear disk, and its interaction with stars and black holes is still lacking to our knowledge. Steps towards this goal could be in our view the pioneering semi-analytical study of \cite{vilkoviski02} or the detailed gas- and stellar dynamical of \cite{johansson09a,johansson09b}, but the latter lack the proper resolution of previously cited purely stellar dynamical work to follow the sub-parsec evolution of the binary SMBH well. One of the most detailed descriptions of physical processes between stars and gas in galactic nuclei has been presented by \cite{ciotti09,ciotti10,shin10a,shin10b}. Their central resolution is much better than that of \cite{johansson09a,johansson09b}, but their models are spherically symmetric and unable to resolve properly the collisional evolution of binary SMBH in dense gas-star systems in galactic nuclei. Another preliminary approach of \cite{just11} follows the star-disk interactions with direct high-accuracy stellar dynamical models (as a follow up to \cite{vilkoviski02}, but still uses a stationary disk model).

In this study we investigate following mechanisms important in determining the structure and evolution of collisional galactic nuclei embedding growing SMBHs 

\begin{itemize}
\item
{\it Destruction of stars by the SMBH.}
A SMBH defines a ``loss cone'' of orbits that pass within its event horizon or tidal disruption sphere, $r\leq r_{\rm t}$. Indeed the existence of a capture sphere is crucial for solutions like Bahcall \& Wolf's, since it precludes the formation of an isothermal ($f\sim e^E$) distribution of velocities which would necessarily have a very high density near the SMBH \citep{peebles72}. Continued loss of stars to the SMBH also implies that no precisely steady-state equilibrium can exist \citep{shapiro77,baumgardt06}; the nucleus  will expand, on a relaxation time scale, due to the effective heat input as stars are destroyed \citep{shapiro77}.

Continued supply of stars to the SMBH requires some mechanism for loss cone re-population. The most widely discussed mechanism is gravitational encounters, which drive a diffusion in energy (E) and angular momentum (J). The latter dominates the loss rate \citep{frank76,ls77}. Roughly speaking, many of the stars dominated by the BH potential (i.e. within its influence sphere) will be deflected into $r_{\rm t}$ in one relaxation time, i.e. the loss rate is roughly $(M_\bullet/m_\star)(t_{\rm rh})^{-1}$. In a collisional nucleus with $M_\bullet \approx 10^6 M_\odot$, this is $\sim 10^6/(10^{10} yr) \approx 10^{-4} yr^{-1}$.

Stellar disruption has direct observational consequences. Tidally disrupted stars produce X- and UV radiation with luminosities of $\sim 10^{44} {\rm ergs}^{-1}$, potentially outshining their host galaxies for a period of days or weeks \citep{kobayashi04,khokhlov96}. A handful of x-ray flaring events have been observed that have the expected signature \citep{komossa04}, and the number of detections is crudely consistent with theoretical estimates of the event rate \citep{wang04}. Tidal flaring events may dominate the x-ray luminosity function of AGN at $L_{\rm X} \lesssim 10^{44} {\rm erg} s^{-1}$ \citep{rees88,milos06}. Compact objects (neutron stars or stellar-mass black holes) can remain intact at much smaller distances from the SMBH; these objects would emit gravitational waves at potentially observable amplitudes before spiralling in and may dominate the event rate for low-frequency gravitational wave interferometers like LISA \citep{hopman06,eilon09}.

That loss cones can be much more quickly refilled in even only slightly axisymmetric or triaxial nuclei had been realized much earlier in the context of tidal accretion of stars onto single black holes \citep{norman83,yu02,merritt04}. It is quite natural after a galaxy merger that the merger remnant is not spherically symmetric and is not completely void of gas, so one would expect frequent mergers of supermassive black holes as well. It seems this is consistent with the cosmological evolution of galaxy and black hole populations \citep{hirschmann10}, and also the LISA gravitational wave community expects frequent coalescences of binary SMBH in the universe \citep{sesana10}. Tidal Disruptions of stars by  binary black holes have recently only been studied by \cite{chen08,chen09,chen10} and \cite{liu09} in the context of X-ray flares. They find, while the total amount of X-ray flares related to binary SMBH may be small (order 10\%), they could show a special behavior in the form of bursts and interruptions of tidal disruptions.

\item
{\it The Bahcall-Wolf mechanism.}

In a collisional nucleus exchange of energy between stars drives the system towards an approximately steady-state distribution of stars around the SMBH in a relaxation time. For a single stellar mass, this is $f(E)\sim |E|^{1/4}$, $\rho\sim r^{-7/4}$ \citep{bahcall77}. Since galaxies with collisional nuclei probably always have $M \lesssim 10^8 M_\odot$ the tidal disruption sphere is more relevant than the Schwarzschildradius. Another condition for the Bahcall-Wolf solution is that $r_{\rm t}$ is much smaller than $r_{\rm hm}$ ($|E_{\rm t}| \gg GM_\bullet/r_{\rm hm}$), which is the case in real nuclei. It represents a 'zero-flux' solution. Nevertheless, in the numerical solutions, the steady-state flux is found to be small but non-zero. The flux is determined by the rate at which stars can diffuse into the disruption sphere at $r_{\rm t}$.

The Bahcall-Wolf solution has been verified in a number of other studies based on fluid \citep{amaro04} or Monte-Carlo \citep{marchant80,freitag02} approximations to the Fokker-Planck equation. And it has been tested via direct NBody integrations, avoiding the approximations of the Fokker-Planck formalism \citep{preto04, baumgardt04}.

Observational measurements of the galactic center reveal a stellar cusp, which appears to be flatter than expected in a collisionally relaxed population around a SMBH \citep{schoedel07,do09,merritt10}. Possible explanations for the absence of a cusp in the observed stars around Sgr A* include mass segregation (which leads to the formation of flatter cusps by lighter stars and steeper cusps by massive central stars, by relaxation), and the destruction of the envelopes of giant stars in the densest parts of the cluster \citep{dale09}. It is not clear that Bahcall-Wolf cusps are present in any other galaxy however, both because relaxation times are generally $\gg 10^{10}$ yr, and also because they are difficult to be observationally resolved.
 
Moreover, if binary SMBHs formed during galaxy mergers, they can destroy dense nuclei, as has been observed in the central density profiles or 'mass deficits' of bright elliptical galaxies \citep{merritt06}. An important question is whether the existence of dense cusps at the centers of galaxies such as the Milky Way and M32 implies that no binary SMBH was ever present, or whether a collisional cusp could have spontaneously regenerated after being destroyed. 
\end{itemize}

The relative importance of these and other various mechanisms, like physical collisions between stars, gas inflow to the SMBH, and the nature of the dark matter that permeate galaxies and their possible interactions, are still not well understood. One of the most advanced physical descriptions including a detailed multi-phase treatment of interstellar matter and mechanical and thermal feedback due to stars and central black holes (AGN) has been presented by \cite{ciotti09,ciotti10} and \cite{shin10}. Their approach, however, still lacks spatial and physical resolution compared to isolated galactic models.

\section[]{The Method}

We define stellar accretion via the loss cone, which in a spherical galaxy is given by orbits, which specific energy ($\epsilon$) and angular momentum ($J$) lie within 
\begin{equation}
J^2 \le J_{\rm lc}^2(\epsilon) \equiv 2r_{\rm t}^2 [\phi(r_{\rm t})-\epsilon] \simeq 2GM_{\bullet} r_{\rm t}
\label{jlc}
\end{equation}

G is the gravitational constant, $\phi$ the BH potential and $J_{\rm lc}$ denotes the loss cone size. Stars of mass $m_\star$ and radius $r_\star$ that come within a distance
\begin{equation}
r_{\rm t} = \alpha \ r_{\star} \left(\frac{2M_{\bullet}}{m_\star}\right)^{1/3}
\label{rtid}
\end{equation}
of the central black hole with mass $M_{\bullet}$ will be tidally disrupted and accreted. 100 \% accretion efficiency is assumed. We include in this definition a free parameter $\alpha$, for scaling of the tidal radius in our simulations (see description in Chapter 3). For $M_{\bullet} \lesssim 10^8 M_\odot$, the tidal radius does not fall below the Schwarzschildradius, if we assume solar type stars \citep{frank76}

Time scales considered here are of the order of the relaxation time, or
\begin{equation}
t_{\rm r} \approx 0.065\ \sigma^3/(G^2 m_{\star}\rho \ln \Lambda)
\label{eq:trelax}
\end{equation}
where $\rho$ is the mass density, $\sigma$ is the 3D velocity dispersion and $\ln\Lambda$ is the Coulomb logarithm \citep{spitzer87}. We adopt $\Lambda = 0.11 N$. \citep{heggie94}.
The relation between the dynamical or crossing time $t_{\rm dyn}$ and $t_{\rm r}$ is given by
\begin{equation}
t_{\rm dyn} \propto \frac{\ln \Lambda}{N} t_{\rm r}
\label{tdyn}
\end{equation}
where $t_{\rm dyn} = r/\sigma$ (r is a characteristic radius of the system, usually $r_{\rm hm}$). Here we point out the strong dependence of $t_{\rm r}$ on $N$. We will specially scale our models in Sec.3.2.

\subsection{Influence and wandering radius}
Motion of stars surrounding the black hole in a certain region are directly influenced by its gravitational field. This region is given by the influence radius of the BH, defined as the radius, where the mass in stars is of the order of $M_{\bullet}$ (for an isothermal sphere $M(< r_{\rm n})= 2 M_{\bullet}$, Merritt 2003). \cite{frank76} define it by
\begin{eqnarray}
r_{\rm n} \equiv \frac{GM_{\bullet}}{\sigma^2}
\label{infrad}
\label{rinf}
\end{eqnarray}
$\sigma$ is here the one-dimensional velocity dispersion. Though both definitions not always agree well, if the mass distribution is known, the first one is easy to determine.

The motion of a heavy particle in a sea of lighter particles can be treated as Brownian motion. As \cite{chatterjee02} point out, for King models with $W_{0}\ge3$ (as used in the present study) equipartition of the black hole with its surrounding core is a valid assumption \citep{merritt07}. The wandering radius can be define as
\begin{eqnarray}
r_{\rm walk} = 0.5 \left(\frac{m_{\star}}{M_{\bullet}}\right)^{1/2} r_{\rm nb}
\label{rwalk}
\end{eqnarray}
where $r_{\rm nb}$ is the NBody unit of length. This empirical relation provides a decent fit for the measured black hole wandering during the whole simulated time in all considered models. This random walk might hinder the formation of a (7/4) cusp, as it stirs up the central region and smears out the eventual over-densities. Superposed on the random walk, it might have a non-vanishing mean velocity varying on much larger timescales. The black hole exerts high frequency oscillations, and has an additional drift, which is observed in the center of mass vector as well \citep{makino87}. We correct this effect by subtracting every time step the center of mass from the black hole position.
\subsection{Loss cone flux}

In a time $t_{\rm r}$, gravitational encounters between stars can exchange orbital energy and angular momentum. Core collapse (shrinking of the core to zero size and infinite density) does not happen in galactic nuclei embedding a SMBH, since the central potential triggers core expansion and the central density drops.

The concept of a loss cone as introduced by \cite{frank76} can be used to identify stars on an orbit penetrating the black holes Roche lobe.  As these loss cone orbits are disrupted within an orbital period, the stellar mass supply must run out immediately if they are not replenished by relaxation.  The loss cone can be defined as an angle like variable giving the half aperture of the black hole as seen from the stars distance.  Trajectories with a smaller aperture have a peribothron $\le r_{\rm t}$ and will be lost.

In a potential dominated by the black hole, taking advantage of the Keplerian velocity profile,
\begin{eqnarray}
 \sigma=\sqrt{\frac{GM_{\bullet}}{r}} 
\label{sigma}
\end{eqnarray}
 and using Eq.~\ref{jlc} for $\epsilon \ll \epsilon_{\rm t}$ \citep{frank76}, the loss cone angle $\theta_{\rm lc} = v_{\rm lc}/\sigma$ evaluates to 
\begin{eqnarray}
\theta_{\rm lc}(r) = \left(\frac{2}{3}\frac{r_{\rm t}}{r}\right)^{1/2}
\end{eqnarray}

In a steady state spherical system, the only process refilling the loss cone is scattering of stars due to distant gravitational encounters. For a general $r^{-\eta}$ density cusp and $\sigma$ given by Eq.~\ref{sigma}, the relaxation time within the black hole sphere of influence becomes 
\begin{eqnarray}
t_{\rm r}=0.338 \frac{M_{\bullet}^{3/2}}{G^{1/2}m_{\star}\rho_{\rm n}\ln \Lambda r_{\rm n}^\eta}r^{\eta-3/2}
\label{trh}
\end{eqnarray}
where $r_{\rm n}$ and $\rho_{\rm n}$ are radius and mass density at the influence radius, and the angular diffusion per orbital period  $\theta_{\rm D} = (t_{\rm dyn}/t_{\rm r})^{1/2}$ turns out 
\begin{eqnarray}
\theta_{\rm D}^2 = 2.960\ r^{3-\eta}r_{\rm n}^\eta M_{\bullet}^{-2}m_{\star}\rho_{\rm n}\ln \Lambda.
\end{eqnarray}

Comparing the two angles, one customarily defines two regimes:  The empty loss cone or diffusive regime for small $r$ where $\theta_{\rm lc}>\theta_{\rm D}$ and the full loss cone or pinhole regime, where stars can move through the loss cone within one orbital period ($\theta_{\rm lc} < \theta_{\rm D}$). \citet{ls77} showed that the flux into the loss cone peaks at $\theta_{\rm D}=\theta_{\rm lc}$, which defines the critical radius
\begin{eqnarray}
r_{\rm crit}=\left(0.225\ M_{\bullet}^2\frac{r_{\rm t}}{m_{\star} \rho_{\rm n}\ln \Lambda}\right)^{\frac{1}{4-\eta}}r_{\rm n}^{\frac{\eta}{\eta-4}} 
\label{rcrit}
\end{eqnarray}
for the assumption of a power law density profile within the black holes radius of influence.  Hence the last relation is valid for $r_{\rm crit}\lesssim r_{\rm hm}$, which holds for most of the $51$ elliptical galaxies in the sample of \cite{wang04}.

\cite{ls77} showed that the integrated number flux per orbital period, can be approximated to match its value at the critical energy,
\begin{eqnarray}
F(E) \sim F_{\rm E}(E_{\rm crit}) |E_{\rm crit}|
\label{eqn:dotn01}
\end{eqnarray}
 where  $F_{\rm E}$ gives the flux of stars at $E = E_{\rm crit}$. A simplified expression of the expected disruption rate for the steady state solution ($\eta$=1.75), can be obtained by applying the scaling of $N(r) \propto r^{5/4}$, and $t_{\rm r} \propto \sigma^3/n(r) \propto r^{17/8}$ in $F_E \propto N(r)/t(r)$, and $E \propto r^{-1}$, and the scaling of $r(t) \propto t^{2/3}$ during self similar expansion \citep{shapiro77}. It results in $\dot{N}_{\rm crit} \sim t^{-1.25}$. 

We use an equivalent expression \citep{frank76}  
\begin{eqnarray}
\dot{N}_{\rm crit} \propto 
 \left.\frac{4\pi r^3\theta_{\rm lc}^2(r)n(r)}{3 \ t_{\rm dyn}(r)}\right|_{r=r_{\rm crit}}
\label{eqn:dotn02}
\end{eqnarray}
for a $\eta=7/4$ power law cusp and $t_{\rm dyn} \simeq \frac{r^{3/2}}{\sqrt{GM_{\bullet}}}$, to derive
\begin{equation}
\dot{N}_{7/4} \propto 6.39 \sqrt{G}ln \Lambda^{5/9}M_{\bullet}^{-11/18}r_{\rm t}^{4/9} m_\star^{10/9} n_0^{14/9} r_0^{49/18}
\label{disr2}
\end{equation}

And in physical units,

\begin{equation*}
\dot{N}_{7/4} \propto \frac{6.89 \times 10^{-6}}{Myr} \ ln \left(\frac{M_{\bullet}}{2m_\star}\right)^{5/9} \left(\frac{r_\star}{R_\odot}\right)^{4/9} \left(\frac{m_\star}{M_\odot}\right)^{26/27} 
\end{equation*}
\begin{equation}
\left(\frac{M_\bullet}{1000 M_\odot}\right)^{-25/24} \left(\frac{n_0}{{\rm pc}^{-3}}\right)^{14/9} \left(\frac{r_0}{\rm pc}\right)^{49/18}
\label{disr3}
\end{equation}

 This equations will help us to properly scale our simulations, as described in the next section.

\section{Simulations and Results}

Evolution of dense stellar systems harboring growing black holes is studied using direct NBody methods with an implemented tidal disruption procedure. The black hole is treated as a heavy particle with an initial mass of $M_{\bullet}/M_{\rm tot}=0.01$. Particle numbers are N = 10 000 to N= 100 000. Simulations were run in parallel with up to 128 processors on the RZG Power 6 Machine in Garching (Munich), which is a facility of the DEISA project; the 40 nodes Kolob GPU-Cluster (Mannheim Germany), the 85 nodes Laohu cluster (Beijing, China) and the GRAPE cluster Titan at ARI-ZAH (Heidelberg, Germany), which has been recently upgraded with GPU cores. 
In our implementation, by using the MPI parallel NBody6++ \citep{spurzem99} and the parallel GPU code $\varphi$GPU \citep{harfst07}, we let the star proceed to its peribothron where the accretion then takes place.  In the NBody6++ code, we take advantage of the neighbor scheme \citep{ahmad73}, and find candidates for the disruption by only searching the black holes neighbors - thus reducing the computational overhead.  These simulations incorporate zero softening \citep{harfst07} and have an energy conservation $\Delta E/ E_{0}<10^{-4}$, even after $10^4$ NBody time units. For the high N (up to 100K) simulations we use the parallel  $\varphi$GPU code, ready for use with GPU clusters which includes a softening parameter of $10^{-5}$. The energy conservation is of the same order as in the NBody6++ code.

In order to challenge the analytic theory of loss cone diffusion by direct NBody simulations, we have to ensure that the simulations probe the dynamics of interest for real galaxies - the empty loss cone regime.  It is convenient to identify the physics at work by its dominating time scale, which in our case demands a separation of the loss cone depletion time $t_{\rm out}=t_{\rm dyn}$ from the loss cone refilling time
$t_{\rm in} \simeq \theta_{\rm lc}^2 t_{\rm r}$.  Hence the condition to have an empty loss cone is
\begin{eqnarray}
t_{\rm dyn}\ll \frac{r_{\rm t}}{r}\ t_{\rm r} < t_{\rm r}.\label{eq:tuneq}
\end{eqnarray}
For a in-depth discussion of $t_{\rm in}$ we like to refer to the gas model studies by \cite{amaro04}.  A straightforward way to satisfy the above relation is by increasing the particle number $N$ following Eq.~\ref{tdyn}. However, as the $O(N^2)$ scaling of direct algorithms  transforms to $O(N^3)$ for relaxation processes, it was a challenge already for our N=100K runs, specially because of the long integration time required for our purposes. Additionally, we highlight that the inclusion of a heavy black hole particle leads to a wide-spread time step distribution with few very short stepped particles close to the black hole.  This hampers scalability and significantly increases the over-all integration time compared to the standard benchmark cases. Thus, it is still very difficult to obtain models of $N \approx 10^6$ on todays general purpose high performance computers, but we expect in a soon future to be able to perform such runs in our new GPU clusters.

Given the limitations on the particle number and to further separate the time scales, we introduce the magnification factor $\alpha$ (Eq.~\ref{rtid}) to vary the tidal radius in our NBody simulations with this free parameter and in order to determine the scaling behavior of the results as a function of $\alpha$, which ranges between 1 to 1000 (or $\sim10^{-5}r_{\rm hm} < r_{\rm t} < \sim10^{-3}r_{\rm hm}$).

It allows us to improve the performance and as well a deeper study of time dependent stellar accretion. Our model NBody units \citep{heggie86} are scaled to $G=1$, $r_{\rm nb}=1 \ pc$, and $m_{\star}=1M_{\odot}$. With this scaling the stellar radius for the single mass runs turns out to be $r_\ast = 2.52 \times 10^{-8} r_{\rm nb}$  the tidal radius then turns out as $r_{\rm t}(0) = 2.52 \times 10^{-7} r_{\rm nb}$ for a mass ratio of 1000:1.  We are neglecting in this study of single-mass systems, the stellar evolution during the whole simulated time. We are presenting evolution of multi-mass systems in a forthcoming publication, where we will fully incorporate the stellar evolution time scale, which becomes short enough to play a significant role.

The black hole is located initially at the center with zero velocities and the mass of tidally disrupted stars is added completely to its mass every accretion event. In order to provide an actual sink in phase space to drive the diffusion, particles need to be removed from the simulation when entering the tidal radius.  The subsequent accretion is bluntly modeled as a perfect inelastic collision with the black hole particle, where the star is fully accreted and linear momentum is conserved.  The "equations of motion'' read
\begin{eqnarray}
M'_{\bullet} &=& M_{\bullet}+m_{\star}\\
\mathbf{r'_{\bullet}} &=& \frac{1}{M_{\bullet}+m_{\star}} \left(M_{\bullet}\mathbf{r_{\bullet}}+m_{\star}\mathbf{r_{\star}}\right)\\
\mathbf{v'_{\bullet}} &=& \frac{1}{M_{\bullet}+m_{\star}} \left(M_{\bullet}\mathbf{v_{\bullet}}+m_{\star}\mathbf{v_{\star}}\right).
\end{eqnarray}

As initial galaxy models, we use rotating King Models \citep{einsel99}, where we added the black hole particle of mass $M_{\bullet}/M_{\rm tot}=0.01$ in the center. For completeness we employ $W_{0}=(3,6)$ and $\omega_{0}=(0.0,0.6,0.9)$ axisymmetric King models. $W_{0}=3$ King models have larger and denser cores than $W_{0}=6$ King models. In the text, we refer to non-axisymmetric models ($\omega_{0}=0.0$) as an approach to spherically symmetric systems, since the first don't have any flattening due to rotation. We use, complementary to our NBody models, Fokker-Planck approximations to our problem, with a seed BH $M_\bullet = 10^{-5}$ and $N=10^8$ to properly define $J_{\rm lc}$, as in Eq.~\ref{jlc}. Here is the stellar radius $r_\star = 2.52 \times 10^{-8} r_{\rm c}$, and thus $r_{\rm t}(0) = 2.52 \times 10^{-7} r_{\rm }$, where $r_{\rm }$ is the scaling radius in our King Models, equal to the core radius. The value of the initial $r_{\rm t}$ corresponds to a mass ratio $M_\bullet/m_\star =1000$. Table~\ref{tab:runs} summarizes the parameters of the performed NBody and Fokker-Planck runs.

\begin{table}
 \centering
 \begin{minipage}{\linewidth}
  \caption{Overview of the performed NBody runs}
\begin{tabular}{|r|r|r|r|r|r|}
     \hline
     Run Identity& $W_{0}$ & $\omega_{0}$& N & $\alpha$ & \\
   \hline\hline
    10KR1a&3&0.0&10000&10&(a)\\
    10KR1b&3&0.0&10000&100&(a)\\
    10KR1c&3&0.0&10000&1000&(a)\\
    10KR2a&3&0.6&10000&1000&(a)\\
    10KR3b&3&0.9&10000&100&(a)\\
    10KR3c&3&0.9&10000&1000&(a)\\
   \hline
    16KR1a&3&0.0&16000&10&(b)\\
    16KR1b&3&0.0&16000&100&(b)\\
    16KR1c&3&0.0&16000&1000&(b)\\
    16KR3c&3&0.9&16000&1000&(b)\\
   \hline
    16KR4c&6&0.0&16000&1000&(b)\\
    16KR5a&6&0.6&16000&10&(c)\\
    16KR5c&6&0.6&16000&1000&(b)\\
    16KR6a&6&0.9&16000&10&(b)\\
    16KR6c&6&0.9&16000&1000&(b)\\
   \hline
    32KR1b&3&0.0&32000&100&(b)\\
    32KR1c&3&0.0&32000&1000&(c)\\
    32KR3c&3&0.9&32000&1000&(b)\\ 
   \hline
    32KR4c&6&0.0&32000&1000&(b)\\
    32KR5c&6&0.6&32000&1000&(b)\\
    32KR6c&6&0.9&32000&1000&(b)\\
   \hline
    64KR1b&3&0.0&64000&100&(b)\\
    64KR1c&3&0.0&64000&1000&(b)\\
    64KR3a&3&0.9&64000&10&(c)\\
    64KR3c&3&0.9&64000&1000&(d)\\
   \hline
    64KR4c&6&0.0&64000&1000&(d)\\
    64KR5c&6&0.6&64000&1000&(d)\\
    64KR6c&6&0.9&64000&1000&(d)\\
   \hline
    100KR1b&3&0.0&100000&100&(c)\\
    100KR1c&3&0.0&100000&1000&(d)\\
    100KR2c&3&0.3&100000&1000&(d)\\
    100KR3c&3&0.9&100000&1000&(d)\\
   \hline
    100KR4c&6&0.0&100000&1000&(d)\\
    100KR5c&6&0.6&100000&1000&(d)\\
    100KR6a&6&0.9&100000&10&(d)\\
    100KR6c&6&0.9&100000&1000&(d)\\
   \hline
    FPKR1&3&0.0&$1\times10^8$&1& \\
    FPKR3&3&0.9&$1\times10^8$&1& \\
    FPKR4&6&0.0&$1\times10^8$&1& \\
    FPKR6&6&0.9&$1\times10^8$&1& \\
   \hline\hline
\end{tabular}\\
Col. (1) is an identifier for the run, Col. (2) is the King parameter of the initial model $W_{0}$, Col (3) is the rotation-parameter in the King-model $\omega_{0}$,  Col. (4) is the initial number of particles, and Col. (5) is tidal radius magnification factor $\alpha$\\
Notes: computer cluster where simulations were performed.\\
(a) NBody6++ in RZG, (b)  $\varphi$GPU in Titan, (c)  $\varphi$GPU in Kolob, (d)  $\varphi$GPU in Laohu (see text for a brief description of the hardware)
\label{tab:runs}
\end{minipage}
\end{table}
\subsection{Cluster evolution}

\begin{figure*}
\begin{minipage}{\textwidth}
\begin{tabular}{c|c}
\includegraphics[width=0.5\textwidth]{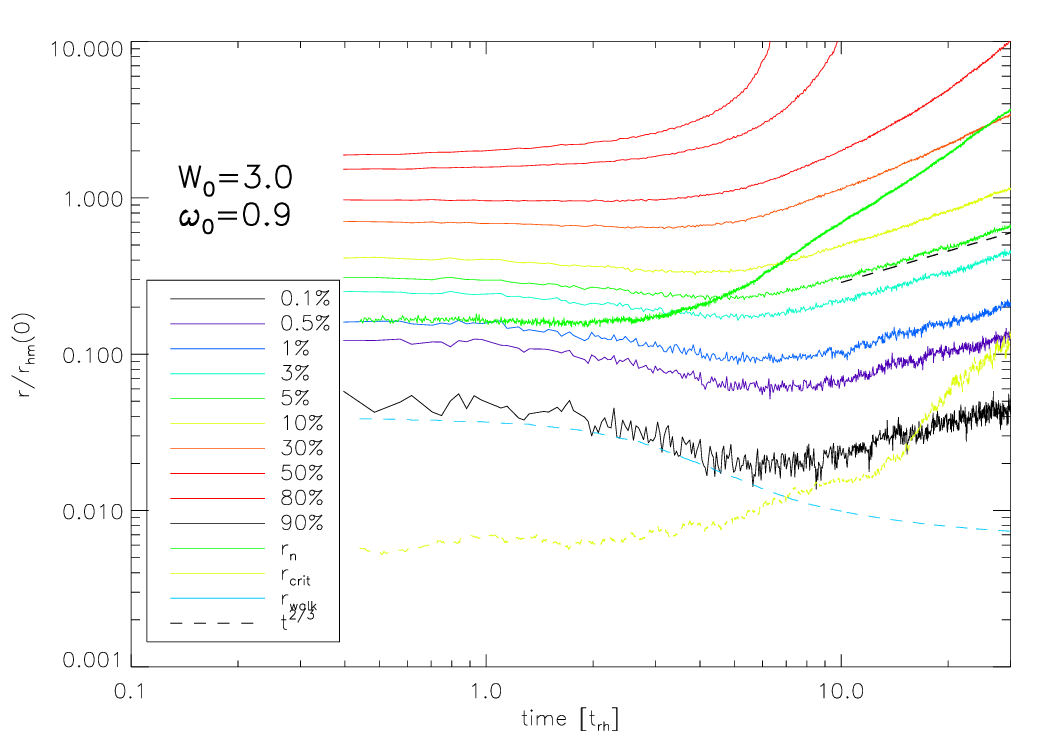} & \includegraphics[width=0.5\textwidth]{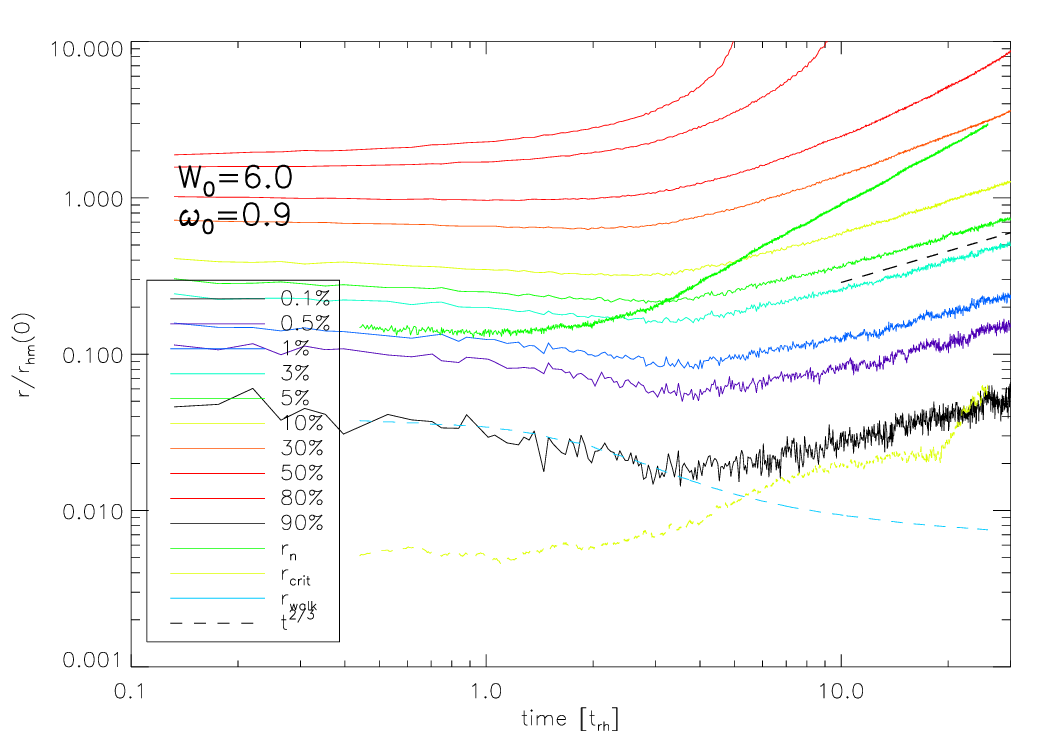}
\end{tabular}
\caption{\small Evolution of the Lagrangian radii for axisymmetric King models 16KR3c (left) and 16KR6c(right). Time is given in units of $t_{\rm rh}(0)$ and radii in units of $r_{\rm hm}(0)$. The system expands after $t_{\rm cc}$ due to the presence of the BH. Evolution of $r_{\rm n}$ (green line), $r_{\rm crit}$ (yellow line),and $r_{\rm walk}$ (light blue line) is shown. The initial $M_{\bullet}/M_{\rm tot}=0.01$}
\label{rlagrange}
\end{minipage}
\end{figure*}

The negative heat capacity of self gravitating systems leads to the process of core collapse.  In the standard picture proposed by \cite{henon65} and \cite{aarseth73}, the singularity is avoided due to the formation and subsequent hardening of tight central binaries. The presence of a star-disrupting black hole can act as an energy source just like binary hardening.  On the event of star-disruption, a bound object is removed from the system, which looses (negative) binding-energy, (energy conservation is established, if the inner energy of the black hole is considered) therefore the system gains energy and expands. \cite{shapiro77} studied the combined effect of star disruption and the change of density in a homological model.  He found a self-similar expansion of the core-radius ($r_{\rm c}$) according to 
\begin{eqnarray}
\frac{r_{\rm c}(t)}{r_{\rm c}(t_{\rm cc})} \propto \left[1+g(M_{\bullet})\ t\right]^{2/3}.
\end{eqnarray}
where $t_{\rm cc}$ is the collapse time \footnote{the collapse time corresponds to the time at which central density grows to infinity, and core radius shrinks to zero. In systems with an energy source, like here, it is the time at which the contraction phase is halted and reversed}, and $g(M_{\bullet})$ also depends on the initial- and on the "minimum-'' core radius and respective density, as well as on the relaxation time. Given the similar underlying physics it is unsurprising that the time-dependence is of the same type ($\propto t^{2/3}$) as in the binary hardening case considered by \cite{henon65} and \cite{goodman84}.

We show the Lagrange radii (radii containing the given percentage of the initial total stellar mass) for the runs 16KR3c and 16KR6c in Figure~\ref{rlagrange}. 
We are using a small particle number, what permits us to obtain longer evolutionary times, in relaxation time scales.

After a less pronounced collapse compared to the case without black hole \citep{kim08}, the Lagrange radii expand self similarly according to the $r \propto t^{2/3}$ law.  This behavior, which has been also seen in gas models \citep{amaro04} and axisymmetric Fokker-Planck models \citep{fiestas10}, is here verified in a self consistent direct NBody simulation for axisymmetric systems.

As we can see, Lagrangian radii give us a qualitative description of the interaction of a growing BH and the stellar mass shells. Initially, Lagrangian radii are dominated by core contraction and the BH mass growth is slow due to the low central density. Later, density grows due to gravitational instabilities (the Lagrangian radii shrink) and the collapse is halted and reversed (mass shells re-expand), while the growing BH potential, dominates the system. The axisymmetric models of Fig.\ref{rlagrange} show a similar evolution as the well studied spherically symmetric case.

In Fig.\ref{rlagrange} we observe that $t_{\rm cc}$ is shorter than the expected value for systems without BH, which are $t_{\rm cc}/t_{\rm rh} \sim 16$ for $W_0=3$ models, and $t_{\rm cc}/t_{\rm rh} \sim 10$ for $W_0=6$ models. This effect is induced by the magnification of the tidal radius and the high initial $M_\bullet$, which enhance stellar accretion, accelerating the reverse of collapse and further expansion. Nonetheless, it does not affect the obtained self-similar expansion. Note that the influence radius approaches the initial $r_{\rm hm}$ during the post-collapse phase. $r_{\rm crit}$ (Eq.~\ref{rcrit}) is also shown in this figures, as well as $r_{\rm walk}$ (Eq.~\ref{rwalk}), which decreases with increasing $M_\bullet$, and becomes more than an order of magnitude smaller than $r_{\rm crit}$, and more than 2 orders of magnitude smaller than $r_{\rm n}$.

Spherically symmetric systems with BH are known to develop steady-state solutions in relaxation time scales. These solutions have characteristic density and velocity dispersion profiles. The density distribution scales theoretically with radius as $r^{-7/4}$ and the velocity dispersion as $r^{-1/2}$. We follow the evolution of our models in this time scale and compare them with the standard theory.

\begin{figure*}
\begin{minipage}{\textwidth}
\begin{tabular}{c|c}
\includegraphics[width=0.5\textwidth]{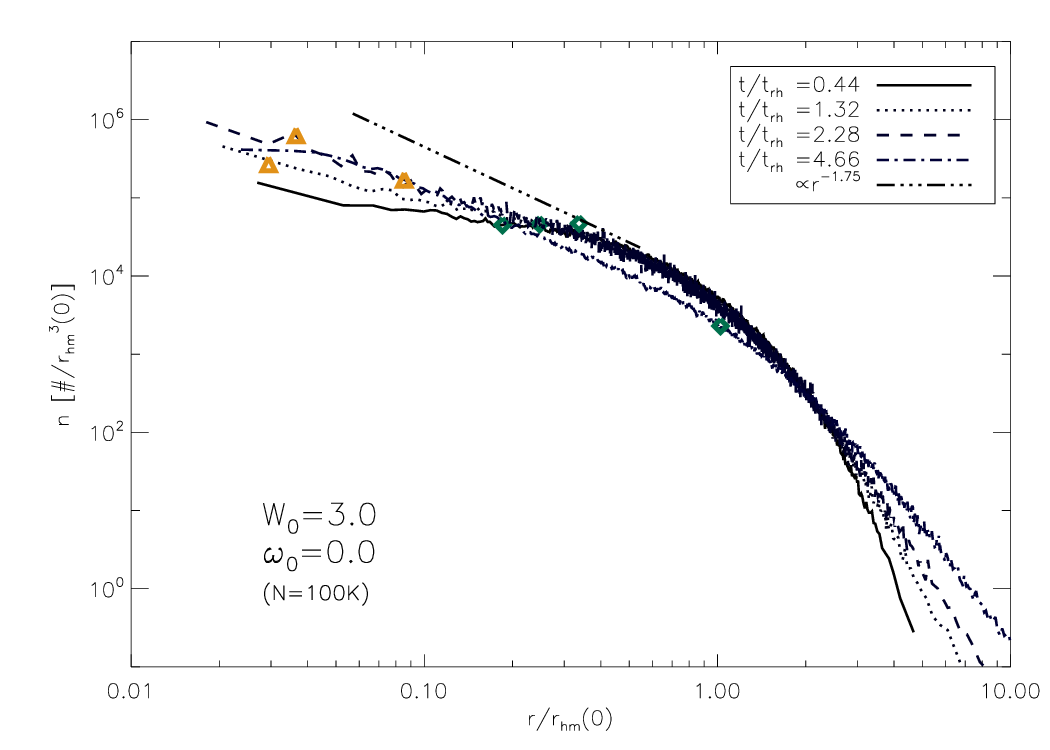}&\includegraphics[width=0.5\textwidth]{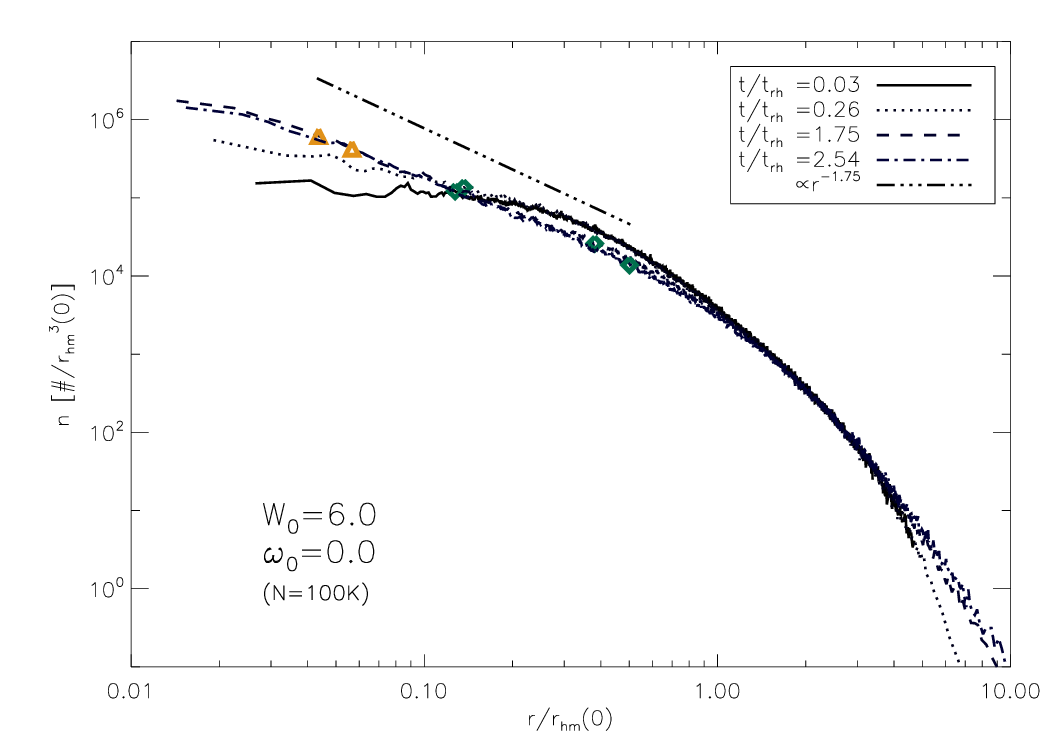}\\
\includegraphics[width=0.5\textwidth]{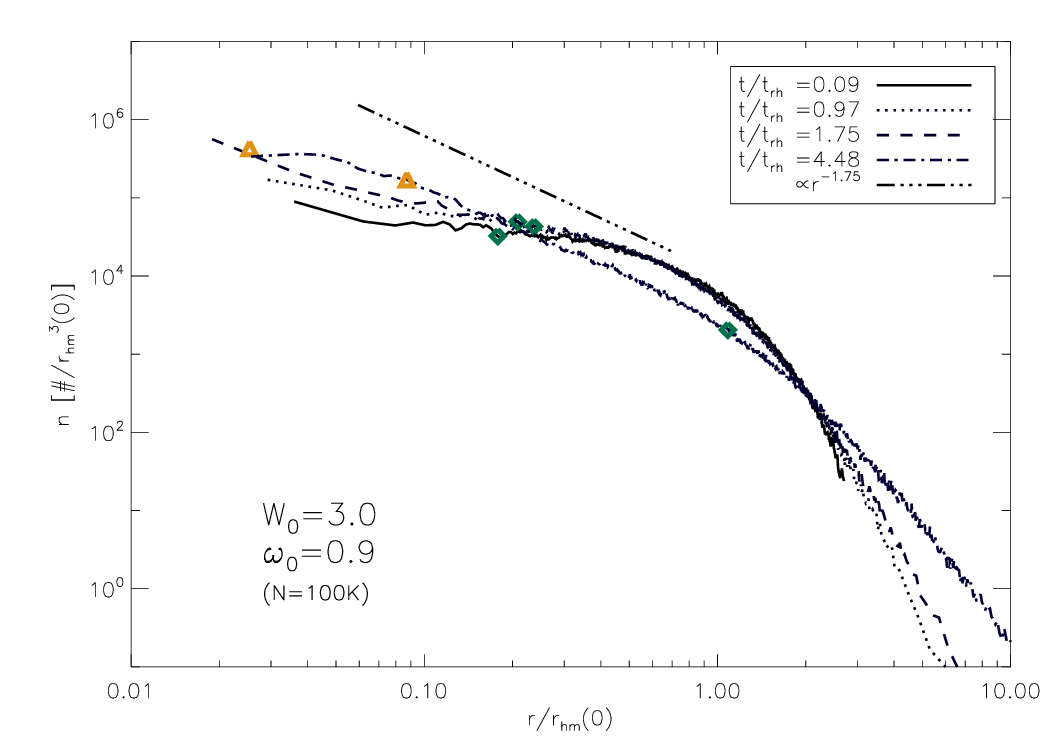}&\includegraphics[width=0.5\textwidth]{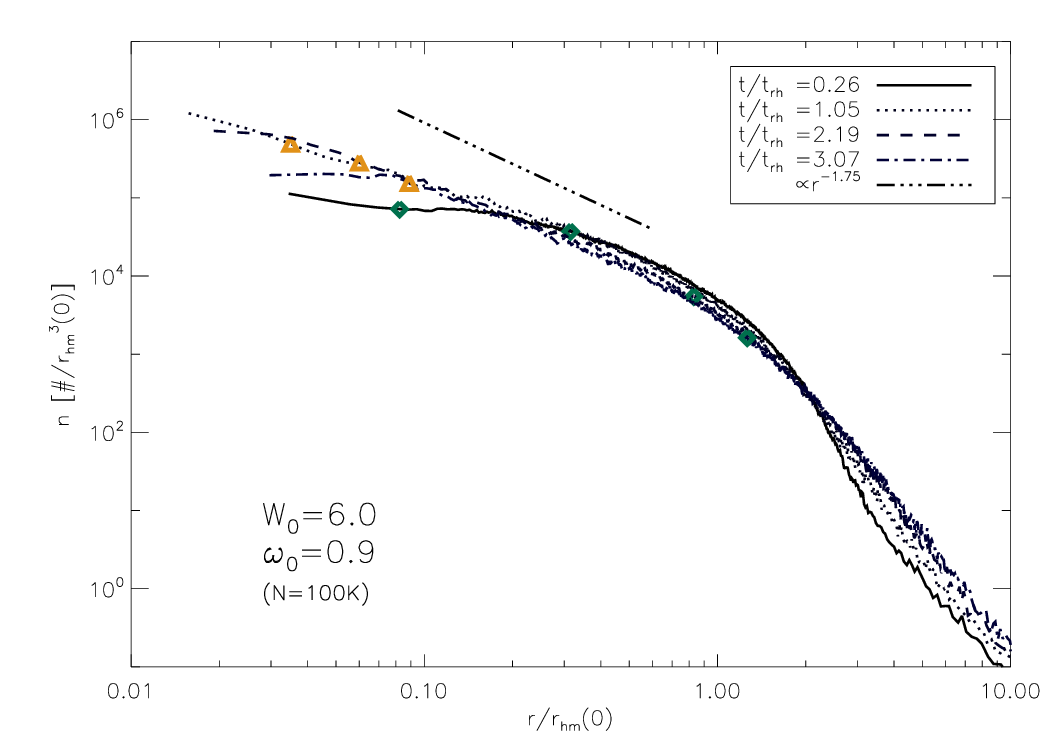}
\end{tabular}
\caption{Evolution of the density profile of non-rotating models 100KR1c and 100KR4c (top), and rotating models 100KR3c and 100KR6 (bottom).  Symbols are: Influence radius $\diamond$ and critical radius $\triangle$. In post-collapse, the radii are well separated and the profile approach the expected zero flux solution between $r_{\rm rcrit}$ and $r_{\rm n}$.}
\label{densevol}
\end{minipage}
\end{figure*}

Fig.~\ref{densevol} shows the evolution of the density profile for the models 100KR1c, 100KR4c (non-rotating), and 100KR3c, 100KR6c (rotating). We use here the highest particle number in order to obtain a better resolution of the cusp. Note that the cores are initially flat and the time needed to build the cusp is of the order of $t_{\rm rh}$. The influence radius (green diamonds) moves outwards while BH mass grows, and the critical radius (orange triangles) appears in our resolution range before the cusp is formed. $r_{\rm walk}$ is smaller than the minimum radius in this figures. In post-collapse, these radii are well separated and the profile approaches the expected zero flux solution between $r_{\rm crit}$ and $r_{n}$.  Note that systems with larger cores ($W_0=3.0$) evolve slower than $W_0=6.0$ models, reaching the latter their final density cusps in shorter times.

During post-collapse the system expands and the cut-off radius in our models extends up to $\sim 10 \ r_{\rm hm}$. Closer to the center, one can see that at later times, the critical radius grows and the cusp is shallower inside this radius, due to the effective stellar accretion around the BH and the growing tidal radius, which perturbs the formation of a cusp at the very center. Although the critical radius is resolved in our simulations, it contains only few dozens of stars. We are performing higher N ($\ge$ 256K) simulations, which will provide a more accurate measurement of the central cusp and specially of the empty loss cone region.

\subsection{Disruption rates}

In our models, stellar accretion is driven by small angle, two-body encounters, which under the influence of the BH gravitational potential, causes that some stars lose energy and move closer to the BH being eventually consumed. During the contraction phase, high central densities are expected to trigger higher BH mass growth rates. Thus, maximal rates occur close to $t_{\rm cc}$, when angular momentum and energy diffusion are most effective.

As previously discussed, in our NBody models, stars in orbits of $J < J_{\rm lc}$, which reach their apocenters inside the tidal radius define an accretion event. In our axisymmetric Fokker-Planck models, stars in orbits of $J_{\rm z} < J_{\rm z,lc}$ are accreted \citep{fiestas10}, being only energy and $J_{\rm z}$ conserved quantities. Since the other angular momentum components are not conserved, accretion can be artificially enhanced in this models, specially during the initial evolution, before self-similar expansion sets on. Regarding initial conditions, our Fokker-Planck models use comparatively smaller BH seeds ($\frac{M_\bullet(0)}{M_{\rm tot}} \sim \ 10^{-5}$), while we fix this value in the NBody models to 0.01.

\begin{figure*}
\begin{minipage}{\textwidth}
\begin{tabular}{c|c}
\includegraphics[width=0.5\textwidth]{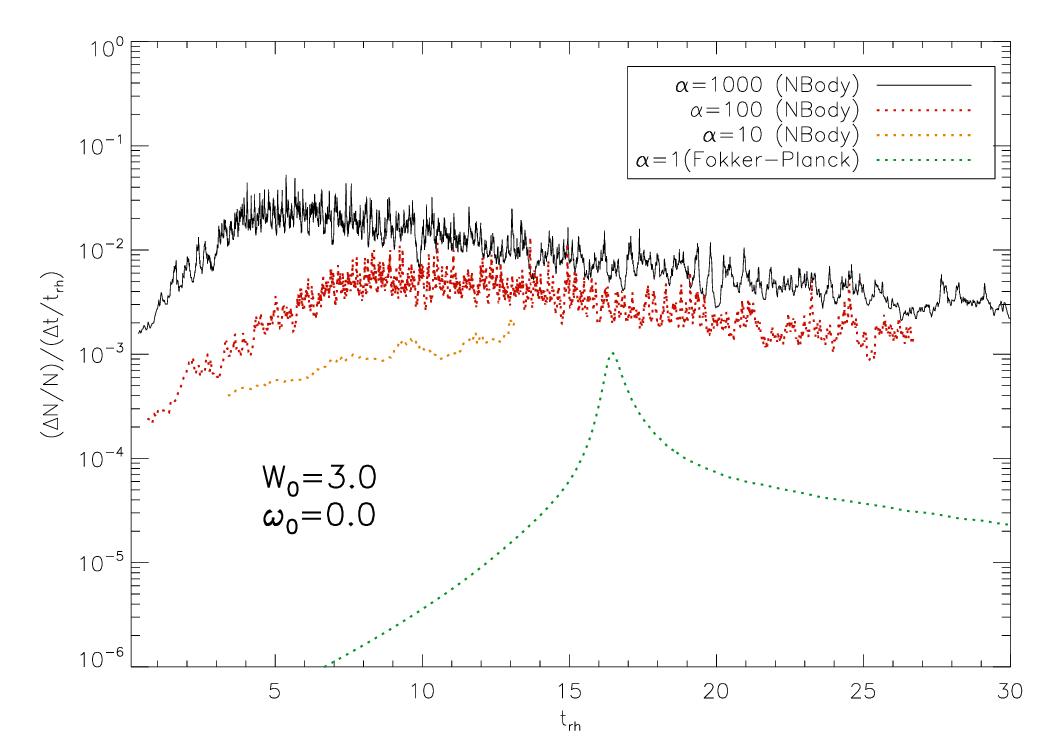} & \includegraphics[width=0.5\textwidth]{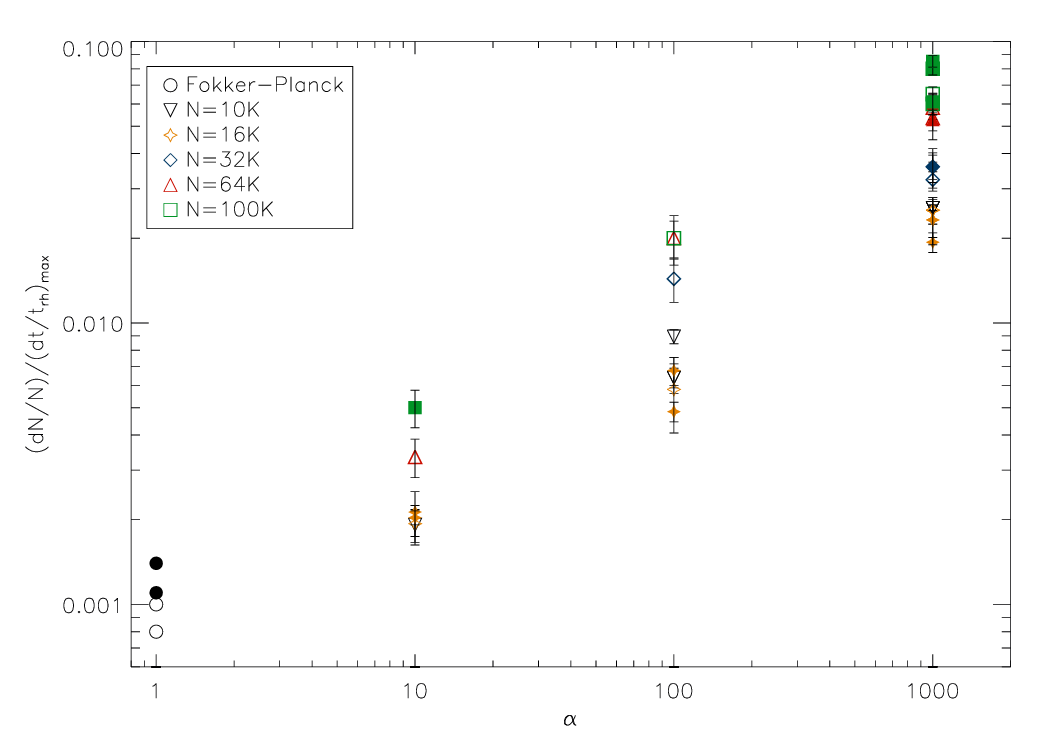}
\end{tabular}
\caption{\small Left: Non-scaled disruption rates for NBody (16KR1a,b,c) and Fokker-Planck models (FPKR1). The influence of the magnified $r_{\rm t}$ in the evolution is clearly seen. The solutions approach the Fokker-Planck results for $\alpha=1$. Right: maximal disruption rates vs. parameter $\alpha$ for all models. Filled symbols are $W_0=6$ models, empty symbols correspond to $W_0=3$ models. The results converge to the ideal case ($\alpha = 1$) as in the Fokker-Planck approximation.}
\label{maxratealpha}
\end{minipage}
\end{figure*}

\begin{figure*}
\begin{minipage}{\textwidth}
\begin{tabular}{c|c}
\includegraphics[width=0.5\textwidth]{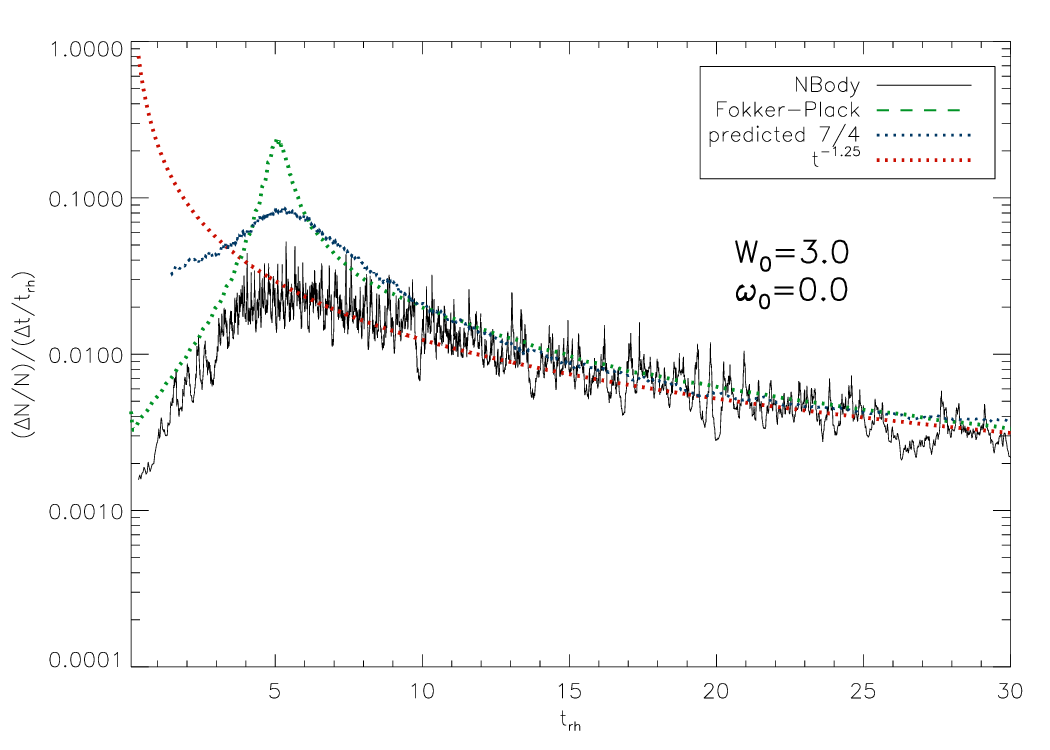}&\includegraphics[width=0.5\textwidth]{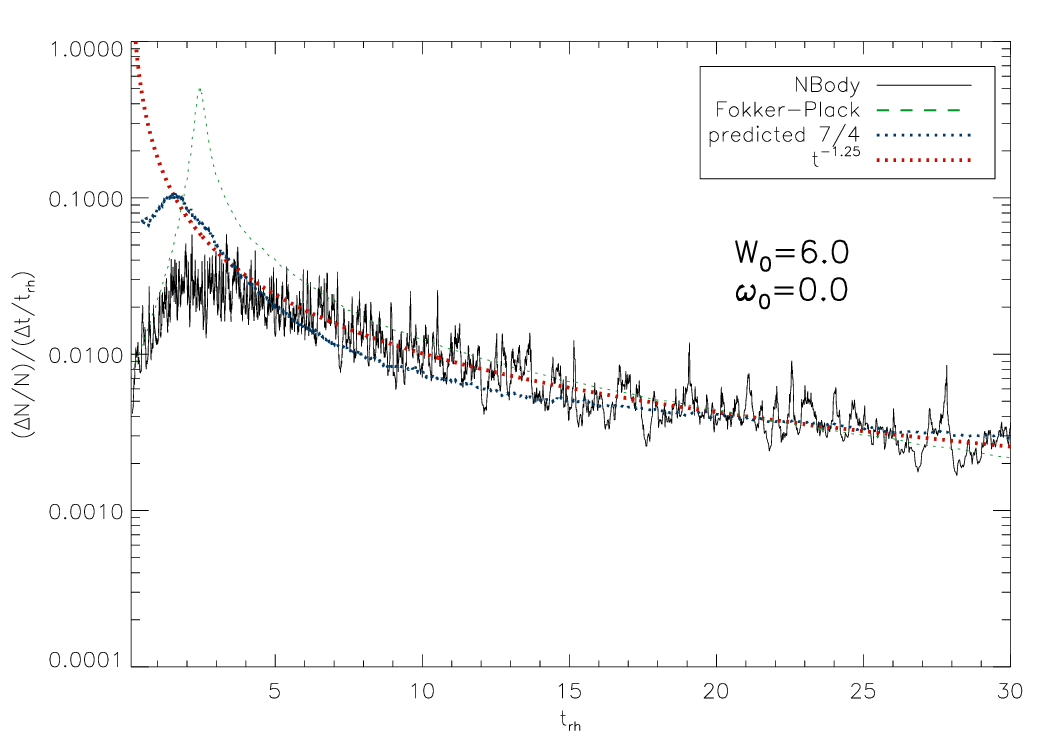}\\
\includegraphics[width=0.5\textwidth]{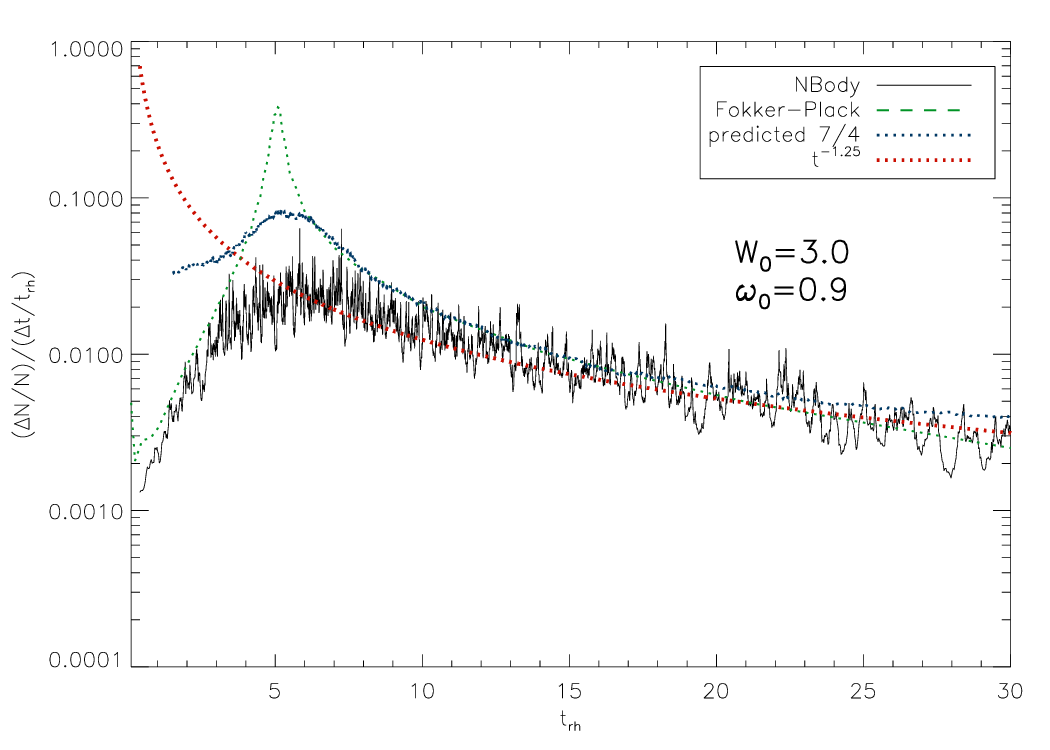}&\includegraphics[width=0.5\textwidth]{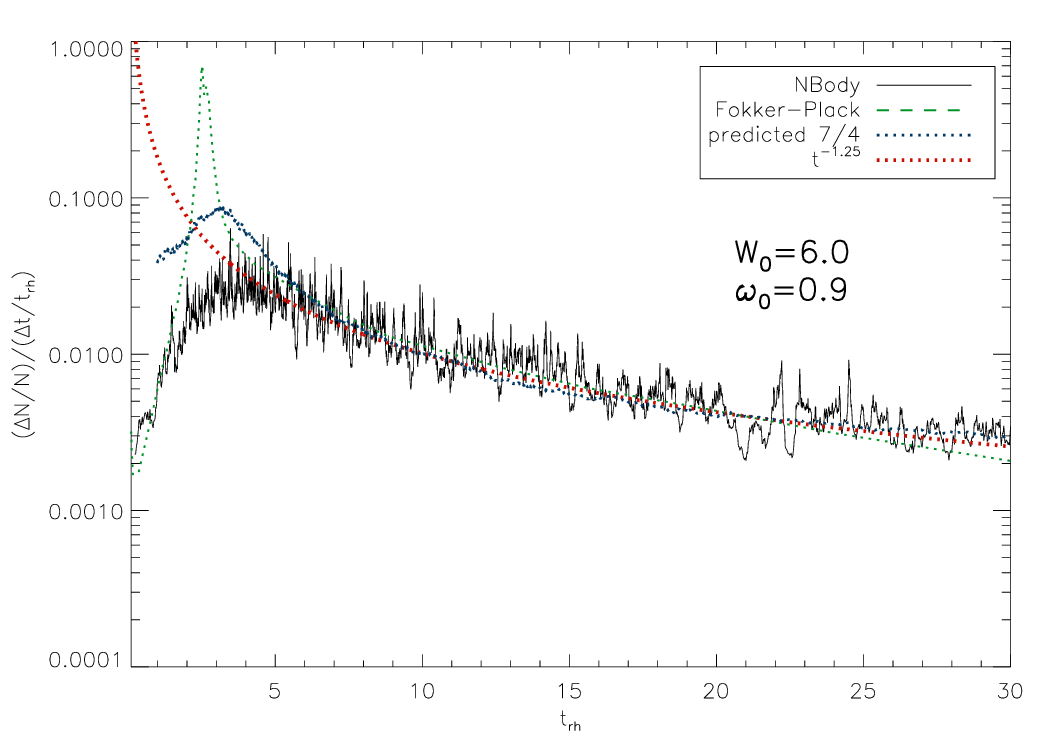}
\end{tabular}
\caption{\small Disruption rates in units of fractional mass per relaxation time for King models 16KR1c, 16KR4 (non-rotating, top) and 16KR3, 16KR6c (rotating, bottom). Nbody results are black lines, green lines are Fokker-Planck models and the expected time dependence for a Bahcall-Wolf cusp ($\propto t^{-1.25}$) is shown by the red line. The predicted rate following Eq.~\ref{disr2} is shown as a blue line. The rate peaks after a few $t_{\rm rh} $ and decreases afterwards. The models follow the analytical predictions in the post-collapse phase.}
\label{disrate}
\end{minipage}
\end{figure*}

Since we expect self-similar evolution during core expansion, independent on initial conditions, as shown in Figs.~\ref{rlagrange} and \ref{densevol}, at this stage both methods should be comparable, if radial and time units are properly re-scaled. We check this by scaling rates, correcting the changes induced in Eq.~\ref{disr2} by $M_\bullet$, $m_\star=1/N$ and $r_{\rm t}$ in each model. First we bring rates to common dimensionless units ($(\Delta N/N)/(\Delta t/ t_{\rm rh})$), and we calculate the factor $\dot{N}_{\rm modelA}/\dot{N}_{\rm modelB}$, by replacing the correspondent values of $M_\bullet$, $m_\star$ and $r_{\rm t}$ of each model. $r_{\rm t}$ enhances the disruption rates by a factor of $\sim$ 1 to 22, (for $\alpha= 1 \ {\rm to} \ 1000$). A variation of N between 10K and 100K modifies the stellar mass $m_\star$ as $1-6 \times 10^{-5}$, and $M_\bullet/m_\star$ changes like $\sim 100 - 1000$. A (-7/4) cusp in the density profile is assumed. We scale finally the results obtained by model B to model A by multiplying the first with the previous factor.

On the other side, as previously discussed, the larger seed BH in the NBody models triggers core heating and enhances accretion events, with the consequence of an earlier bounce of core density. It modifies the time at which density reaches its maximum, and maximal disruption rates appear. We can observe this effect in the evolution of Lagrangian radii (Fig.~\ref{rlagrange}). Thus, self-similar expansion during post-collapse can be compared by bringing the collapse times of the compared models together. 

Fig.~\ref{maxratealpha},left shows disruption rates before scaling. Note that peak rates are higher for higher $\alpha$ and NBody models approach Fokker-Planck models by decreasing this parameter (the 'real' tidal radius is given by $\alpha=1$). Since maximal disruption rates dominate the stellar accretion events over time, they can give us the extent of the influence of the different initial parameters and kinematics (rotation) used in our models. Figure~\ref{maxratealpha},right shows maximal disruption rates ($\dot{N}^{\rm max}$) obtained for all models in dimensionless units $(dM/M_{\rm tot})/(dt/t_{\rm rh})$  against the parameter $\alpha$, as used in the simulations. This plot gives only approximately the $\dot{N} (\alpha)$-dependence, since the $m_\star$-, and $M_\bullet$-dependence are also present. Additionally, the different peak rates at constant $\alpha$, are a consequence of the better resolution reached in high-N Nbody runs, which show higher peaks. Moreover, $\dot{N}^{\rm max}$, for smaller values of $\alpha$ converges to rates corresponding to $\alpha = 1$ (actual $r_{\rm t}$), as we show by plotting the maximal rates of our Fokker-Planck models (black dots in Figure~\ref{maxratealpha}, right).  The measured values of $\dot{N}^{\rm max}$ are shown in Table~\ref{results}. Mainly models using $\alpha=1000$ and $\omega_0=0.0,0.9$ were here selected for comparison. Table~\ref{results} shows as well $\dot{N}^{\rm max}_{\rm scaled}$ scaled to $\alpha$ = 1, as in our Fokker-Planck models. We correct for the $ \dot{N} \propto r_{\rm t}^{4/9}$-dependence, as in Eq.~\ref{disr2}.
 
Fig.~\ref{disrate} shows the disruption rates for models of King Parameter $W_0=3.0$ (left) and $W_0=6.0$ (right). In each figure Nbody results (black lines) together with scaled Fokker-Planck approximations (green lines) are plotted. Additionally the expected time dependence of $\dot{N} \propto t^{-1.25}$, for $\eta=1.75$, following Eq.~\ref{eqn:dotn01}, is shown (red line). We introduce as well the expected rate in our simulations for a (7/4)-cusp by using Eq.~\ref{disr2} (blue line). Rates are averaged over 15 events and smoothed over 20 $t_{\rm nb}$. Rates initially increase fast and reach their maxima close to $t_{\rm cc}$, decreasing afterwards during the expansion of the system. From these results, we can conclude that in the post-collapse phase, our NBody models follow the predicted evolution and agree with the Fokker-Planck results during self-similar expansion. We can further test both models by comparing scaled maximal disruption rates, as shown in column 5 of Table~\ref{results}. In physical units, for a galaxy core of $M \approx 10^9 M_\odot$ and a relaxation time of 10 Gyr, we obtain $\dot{N}_{\rm max} = 1.3 \pm 0.2 \times 10^{-4} M_\odot {\rm yr}^{-1}$. Here we average the results of our ($W_0=0.6, \omega_0=0.0$)-models and correct the $r_{\rm t}$ ($\alpha$ and N)-dependence as in Eq.~\ref{disr2} and Eq.~\ref{rtid}. The corresponding Fokker-Planck models give $\dot{N}_{\rm max} = 1.1 \times 10^{-4} M_\odot {\rm yr}^{-1}$, in agreement with the NBody results. Thus, we can conclude that we obtaine a non-trivial result through all the post-collapse, where NBody and Fokker-Planck models agree with each other, which is of particulary importance, since both methods use an equivalent but not identical treatment of accretion and because definition of relaxation time (our Eq.~\ref{trh}) in galactic nuclei can be eventually over-estimated \citep{madigan11}. This means that we can from these data make predictions for the star accretion rates and other parameters of dry galactic nuclei, which are independent of the previous history. And it shows, that relaxation processes might play an important role in the growth of SMBH. 

\begin{figure}
\includegraphics[width=0.5\textwidth]{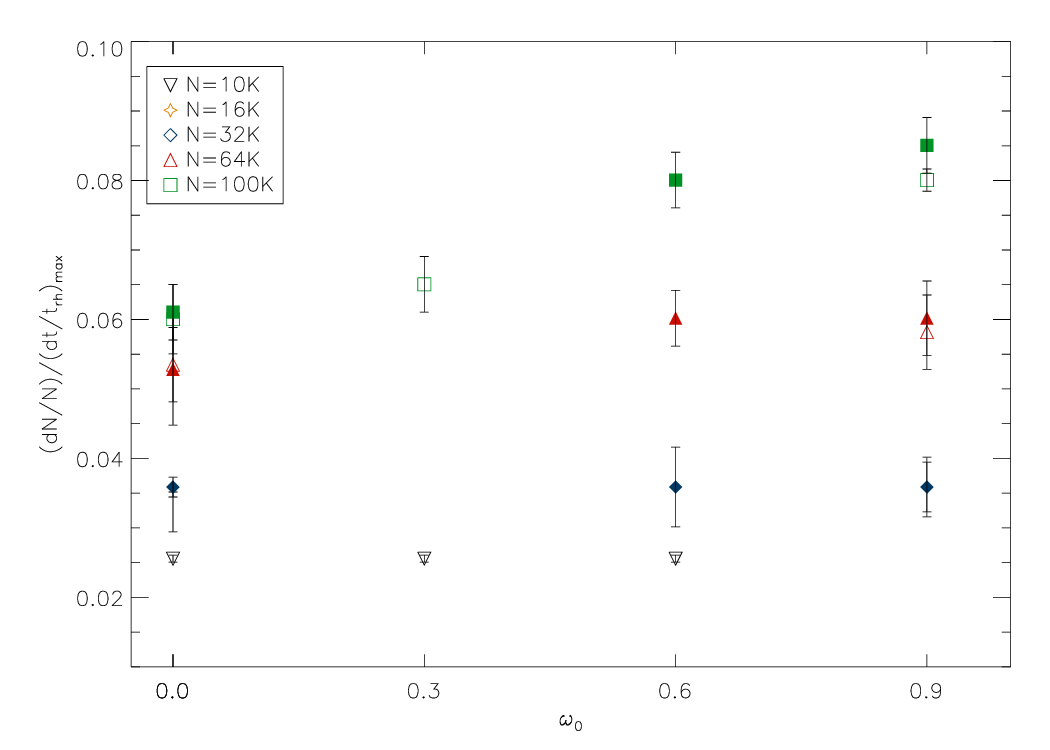}
\includegraphics[width=0.5\textwidth]{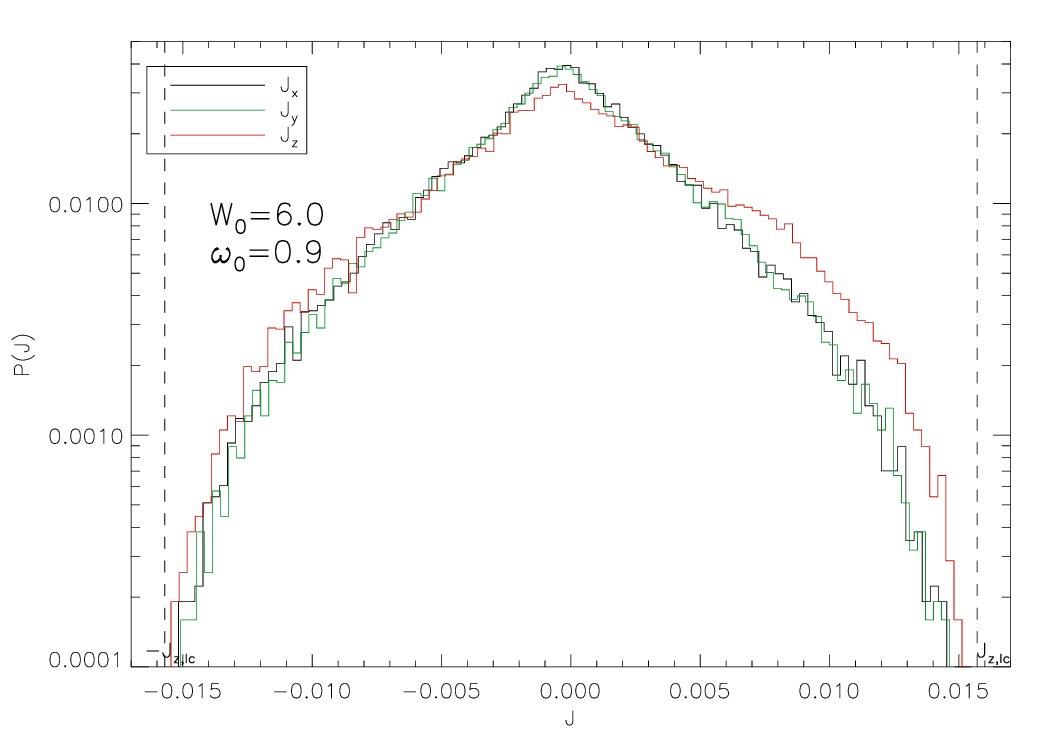}
\includegraphics[width=0.5\textwidth]{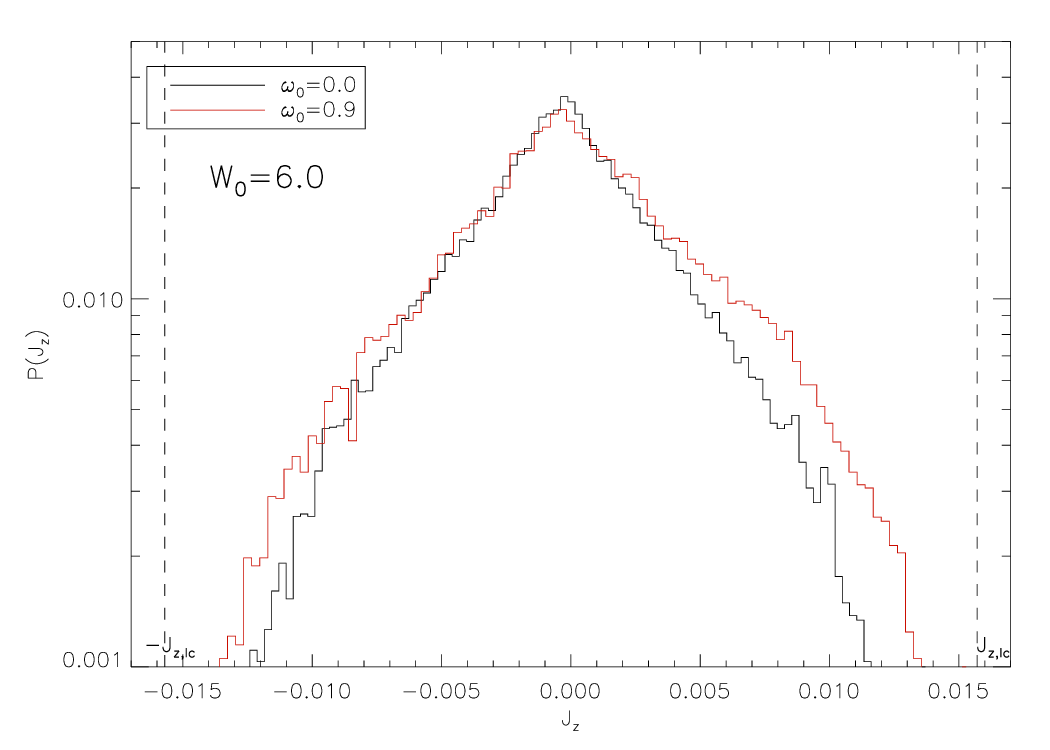}
\caption{\small Top: Maximal disruption rates vs. rotational parameter $\omega_0$ for all models. Symbols are like in Fig.~\ref{maxratealpha}, right. Specially high N runs show how rotation influences the disruption rates. We keep a constant parameter $\alpha=1000$ for comparison. Middle: Distribution of $J_{\rm x}, J_{\rm y}, J_{\rm z}$ of accreted stars for the rotating model 100KR6. A higher fraction of prograde rotating stars is accreted. Bottom: Distribution of $J_{\rm z}$ of accreted stars for the models 100KR4 (non-rotating) and 100KR6 (rotating). The excess of accreted prograde rotating stars leads to the obtained higher masses. The maximum $J_{\rm z,lc}$, for the largest energy of the stellar orbits, is indicated in the middle and bottom panels with a vertical dashed line.}
\label{maxrate}
\end{figure}

By comparing non-rotating with rotating models in Table~\ref{results}, column 4, we can observe that the latter show slightly higher peak rates, for constant N. Maximal rates vs the rotational parameter $\omega_0$ for all models are shown in Fig.~\ref{maxrate}, top. Low N runs show roughly constant peaks, while the better spatial resolution reached by high N runs, specially at $r_{\rm crit}$ (as seen in Fig.~\ref{densevol}), show higher disruption rates for higher $\omega_0$. The higher peaks reached by rotating systems lead to the BH masses shown in Table~\ref{results}. Final BH masses ($M_{\bullet,f}$), are measured at times, which are kept the same for constant N and $W_0$ to facilitate the comparison between the models. We observe specially in the N=100K runs, that non-rotating models reach a similar final mass, since rotating models converge to a $\sim 20 \%$ higher mass, independent of $W_0$. It implies a direct influence of rotation in the process of stellar accretion.

In order to understand how this effect is triggered, Fig.~\ref{maxrate},middle shows the distribution of $J_{\rm x},J_{\rm y},J_{\rm z}$ of accreted stars for the model 100KR6 (rotating). Values are fractions of the total number of accreted stars. All distributions peak at $J=0$, as expected. The contribution of $J_{\rm x},J_{\rm y}$ to accretion is symmetric with respect to $J=0$, while a higher fraction of prograde rotating stars ($J_{\rm z}>0$) is consumed by the BH. This was expected, since our initial models rotate in the positive direction of $J_{\rm z}$. Fig.~\ref{maxrate},bottom shows the distribution of $J_{\rm z}$ of accreted stars for the model 100KR6 (rotating) in comparison to the non-rotating model (100KR4). We observe that rotating models show an excess of accreted prograde rotating stars.

\begin{figure}
\includegraphics[width=0.5\textwidth]{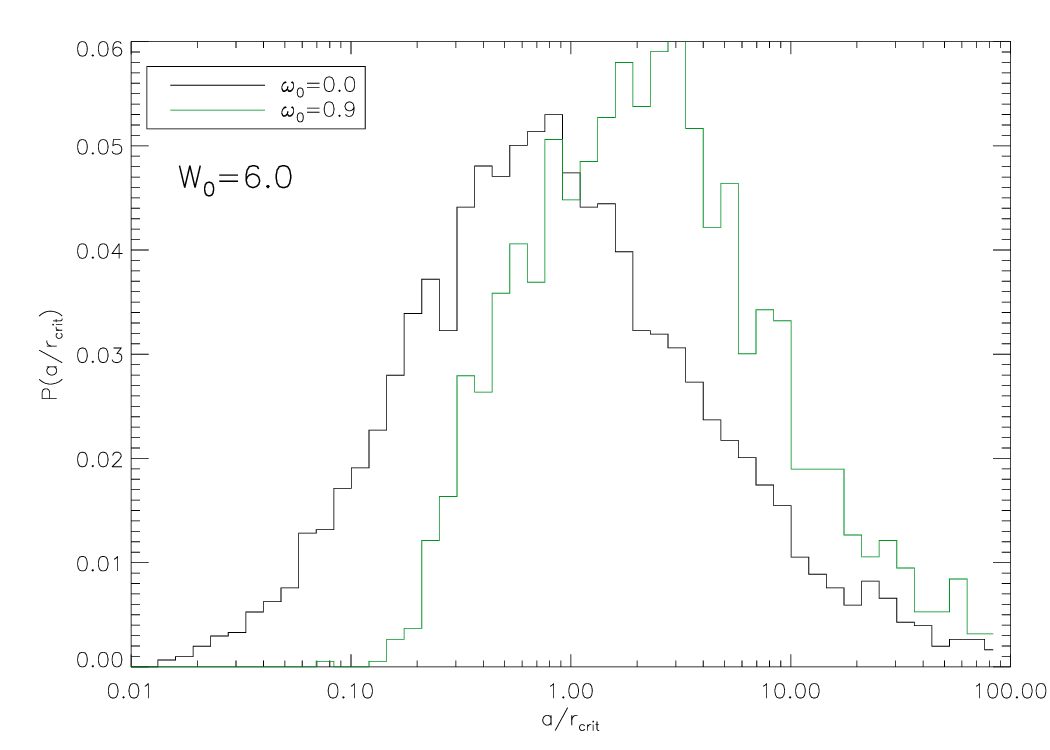}\\
\includegraphics[width=0.5\textwidth]{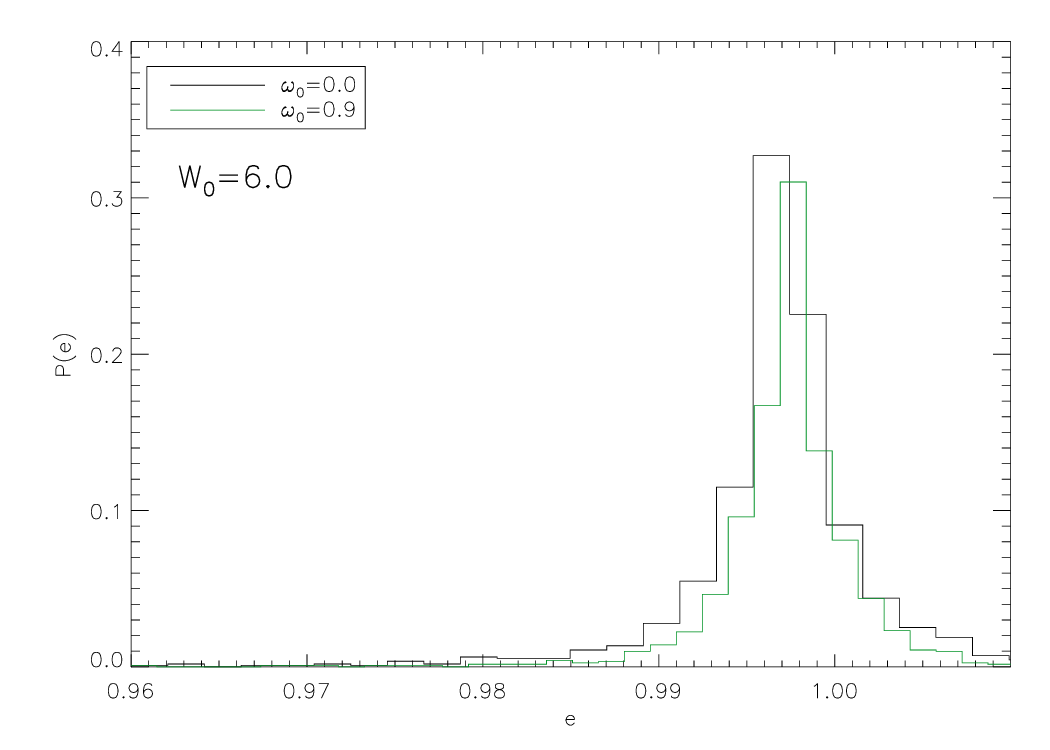}
\caption{\small Top: Histogram of semimajor axis over critical radius of orbits of accreted stars in models 100KR4c (non-rotating) and 100KR6c (rotating). Most accreted stars in the rotating models have semi-major axis few times larger than $r_{\rm crit}$. Distribution of accreted stars in the non-rotating model peaks at $a/r_{\rm crit}=1$ as expected. Only bound orbits are shown in this figure.  Bottom: Histogram of eccentricities for the same accreted stars. Distribution in both models is similar. A contribution of unbound orbits is present. }
\label{hist}
\end{figure}

We show in Fig.~\ref{hist} the radial distribution of the semimajor axis, normalized by $r_{\rm crit}$, of accreted stars in models 100KR4c (non-rotating) and 100KR6c (rotating). A larger fraction of accreted stars in the rotating models have semi-major axis few times larger than $r_{\rm crit}$. Distribution of accreted stars in the non-rotating model peaks at $a/r_{\rm crit} \sim 1$ as expected from the loss-cone theory \citep{ls77}. The distribution of eccentricities is similar in both models (Fig.~\ref{hist}, bottom). It shows a steep maximum above $e=0.995$.
From Figs.~\ref{maxrate} and \ref{hist} we can conclude, that the excess of accreted rotating stars, origins mainly from regions outside $r_{\rm crit}$. According to loss cone theory, diffusion of energy and angular momentum is higher than the loss cone size in this region (full loss cone regime), and orbits in this region are able to escape the loss cone in a dynamical time. In order to obtain a larger $r_{\rm crit}$, as in our rotating models, $t_{\rm in} \sim \theta_{\rm lc}^2 t_{\rm r}$ should be larger, or $t_{\rm out} \sim \theta_{\rm d}^2 t_{\rm r}$ shorter in these region \footnote{the crossing point in Fig.13 of \cite{amaro04} is shifted to the right in our rotating models, wenn compared to non-rotating models}. This hints to a breakdown of the classical theory which depends on the conservation of angular momentum that is not given in the rotating models. An enhanced relaxation in a system that is supported by bulk rotation, as compared to a non-rotating one which is pressure supported, could explain this effect,\citep{atha01}. A decrease in the velocity dispersion for $r < r_{\rm n}$, which could modify substantially $t_{\rm r} \propto \sigma^3$ within the BH influence zone, was nevertheless not detected in our models. It is thus tempting to interpret this behavior in terms of the additional presence of orbits with non-conserved $J_{\rm x}$,$J_{\rm y}$ angular momentum (e.g. box orbits). To further investigate this effect, an orbit study of the accreted stars should be performed, which we must leave open for future research.

\begin{table}
 \centering
 \begin{minipage}{\linewidth}
  \caption{Results of the simulations}
\begin{tabular}{|r|r|r|r|r|r|r|}
     \hline
     Identity& $W_{0}$ & $\omega_{0}$& $\dot{N}^{\rm max}$ & $\dot{N}^{\rm max}_{\rm scaled}$ & $M_{\bullet,f}$&$t/t_{\rm rh,f}$\\
   \hline\hline
    16KR1c&3.0&0.0&.0252&.0008&0.2804&31.5\\
    16KR3c&3.0&0.9&.0254&.0008&0.2788&31.5\\
   \hline
    16KR4c&6.0&0.0&.0193&.0008&0.2836&31.5\\
    16KR5c&6.0&0.6&.0231&.0010&0.2803&31.5\\
    16KR6c&6.0&0.9&.0232&.0010&0.2785&31.5\\
   \hline
    32KR1c&3.0&0.0&.0323&.0012&0.2048&10.2\\
    32KR3c&3.0&0.9&.0358&.0012&0.2102&10.2\\
   \hline
    32KR4c&6.0&0.0&.0358&.0013&0.2624&13.0\\
    32KR5c&6.0&0.6&.0359&.0013&0.2776&13.0\\
    32KR6c&6.0&0.9&.0358&.0013&0.2799&13.0\\
   \hline
    64KR1c&3.0&0.0&.0535&.0018&0.1592&5.5\\
    64KR3c&3.0&0.9&.0468&.0016&0.1791&5.5\\
   \hline
    64KR4c&6.0&0.0&.0528&.0017&0.2283&6.6\\
    64KR6c&6.0&0.9&.0601&.0020&0.2647&6.6\\
   \hline
    100KR1c&3.0&0.0&.0600&.0018&0.2478&6.9\\
    100KR3c&3.0&0.9&.0800&.0020&0.2785&6.9\\
   \hline
    100KR4c&6.0&0.0&.0611&.0019&0.2542&6.9\\
    100KR6c&6.0&0.9&.0850&.0026&0.3175&6.9\\
   \hline
   FPKR1&3.0&0.0&.0008&.0008&0.0011&30\\
   FPKR3&3.0&0.9&.0011&.0011&0.0012&10\\
   FPKR4&6.0&0.0&.0011&.0011&0.0010&30\\
   FPKR6&6.0&0.9&.0017&.0017&0.0012&10\\
   \hline\hline
\end{tabular}\\
Col. (1) is the identifier for the run, Col. (2) the King parameter of the initial model $W_{0}$, Col (3) the rotation-parameter in the King-model $\omega_{0}$, Col (4) are the maximal disruption rates in units of $(dM_\bullet/M_{\rm tot})/(dt/t_{\rm rh})$ , Col (5) are the rates scaled to $\alpha=1$, Col. (6) is the final black hole mass, and Col (7) is the time in relaxation time units at which the final mass was measured.
\label{results}
\end{minipage}
\end{table}

\subsection{Rotational velocity}

\begin{figure}
\includegraphics[width=0.5\textwidth]{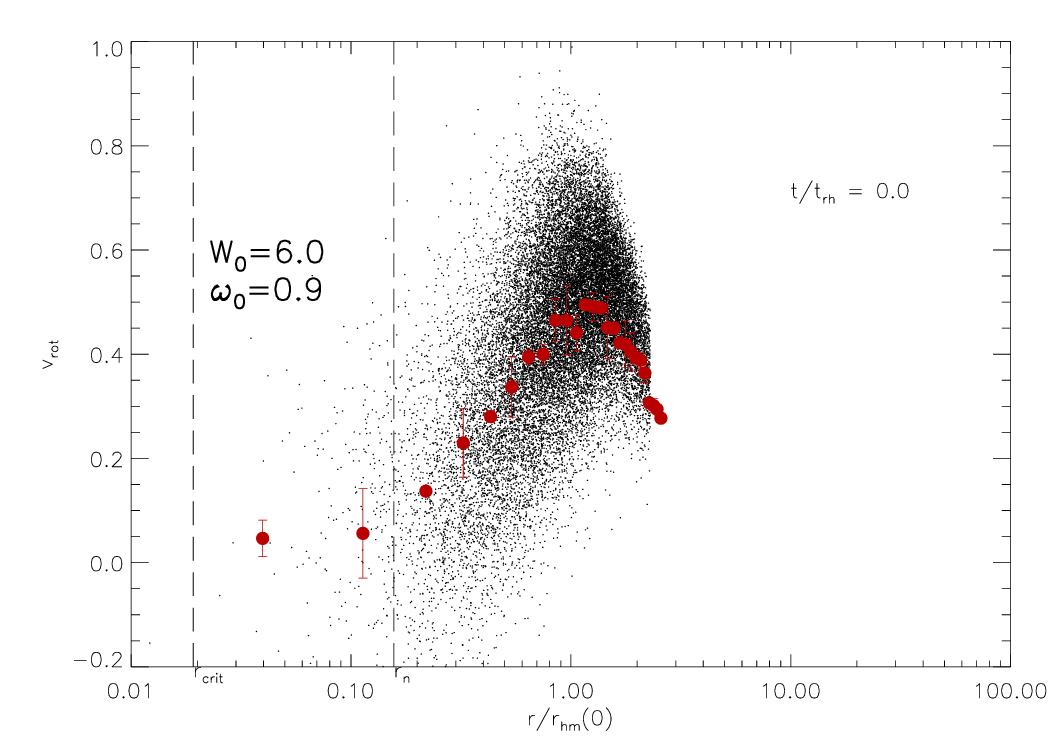}\\
\includegraphics[width=0.5\textwidth]{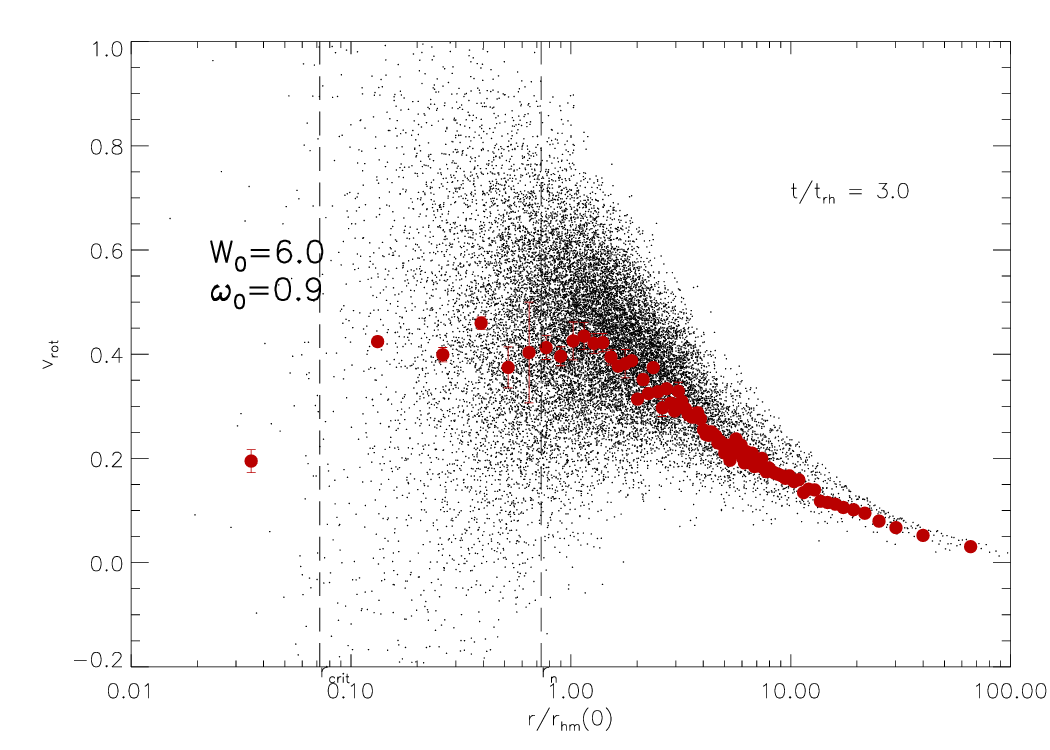}
\includegraphics[width=0.5\textwidth]{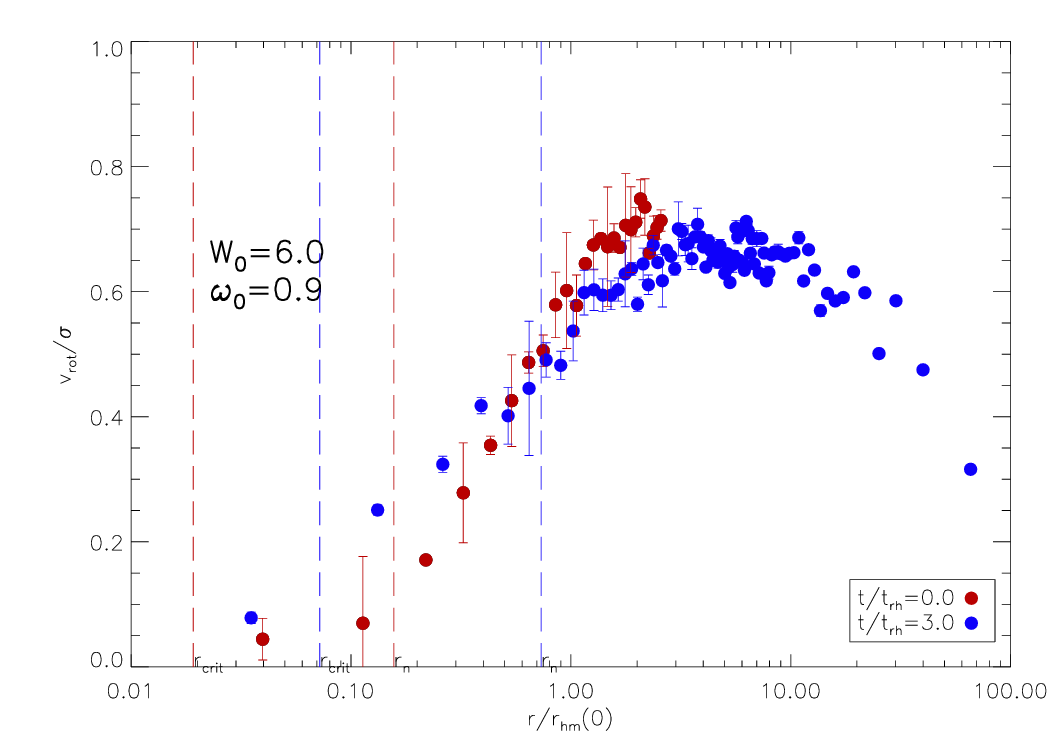}
\caption{\small Top: Initial distribution of rotational velocity of stars in the rotating model 100KR6c. Rotational velocities are averaged in bins of 5 stars and over 5 NBody time units. Vertical dashed lines mark $r_{\rm crit}$ and $r_{\rm n}$ in units of $r_{\rm hm}$ at the given time. The averaged profile is overplotted (red dots) and error bars are included. Middle: rotational velocity of stars after relaxation. A rotating core can be detected for $ r_{\rm crit} \lesssim r \lesssim r_{\rm n}$. Bottom: Initial and final rotational velocity over velocity dispersion for the same model. Growth of $v_{\rm rot}/\sigma$ inside $r_{\rm n}$ is less pronounced because of the central cusp of $\sigma$.}
\label{vrot}
\end{figure}

In Fig~\ref{vrot},top we show the initial distribution of the velocity component correspondent to $J_{\rm z}$, of stars for the model 100KR6c. The initial rotating King Models show a maximum of rotation ($v_{\rm rot,max}$) at around $r_{\rm hm}$ and central rigid body rotation. We plot velocities of bins of 5 stars, in order to get a more detailed stellar distribution, and average over 5 NBody time units. Vertical dashed lines mark the radius of influence of the BH and $r_{\rm crit}$ in units of $r_{\rm hm}$ at the given time, as indicated. The rotation profile from bins of 50 stars is overplotted (red dots). The evolved distribution of rotational velocities for the system after relaxation ($\sim 3 \ t_{\rm rh}$) is shown in Fig.~\ref{vrot}, middle. As a consequence of angular momentum transport through gravitational scatterings, the original $v_{\rm rot,max}$ decreases during the evolution. In the model $v_{\rm rot,max}(t_{\rm cc})/v_{\rm rot,max}(0) \sim 0.85$. Additionally, one can observe that  $v_{\rm rot,max}$ moves inwards, in a region close and inside the influence radius, leading to an increasing of rotation in this region, with respect to the initial configurations. With constant $\omega = 0.9$, as an initial rotational parameter, this effect is more pronounced in concentrated models (100KR6c, $W_0 = 0.6$) than in models with larger cores (100KR3c, $W_0 = 0.3$), because the first contain initially more amount of rotational energy with respect of kinetic energy ($T_{\rm rot}/T_{\rm kin} \sim 0.3$) than the second ones ($T_{\rm rot}/T_{\rm kin} \sim 0.1$).  After the system relaxes, central stars rotate with velocities in average larger than the original maximum, initially located in the outer parts of the system.

Consider the region $r_{\rm crit} < r < r_{\rm n}$, where stars are dominated by the BH central potential. Stars populate always more this region in time (Fig.~\ref{vrot},middle),  while they interact dynamically with other stars. A fraction of them will be disrupted, when their orbits reach $r < r_{\rm t}$. As we discussed, the excess of disrupted stars in rotating models comes mainly from this region, and is dominated by stars in orbits with positive $J_{\rm z}$. It is not surprising, since the concentration of rotating stars in the BH zone of influence is triggered by the initial configuration in our axysimmetric systems. Moreover, the growing central density, caused by gravitational and gravogyro instabilities \citep{einsel99} together with angular momentum transport, enhances this effect, specially before expansion sets on.

The few counter-rotating stars observed in Figure~\ref{vrot}, middle, are remaining stars from the initial distribution, which together with pro-rotating stars lead to no rotation in the very center. The enhancement of rotation inside the influence radius builds a wide maximum of rotating stars inside $r_{\rm n}$, which is now close to $r_{\rm hm}$. At a radius inside $\sim r_{\rm crit}$ one finds only a few dozens of stars (in our N=100K runs), and is difficult to talk about rotation in this region. Moreover, in the region $r_{\rm crit} \lesssim r \lesssim r_{\rm n}$ a rotating core can be detected. By comparison with Fig.~\ref{densevol}, we observe that this is the region where the cusp forms, and the BH potential dominates the stellar environment.

In order to investigate how strong rotation dominates the dynamics at this evolutionary stage, we show the rate of rotational velocity vs. one dimensional velocity dispersion ($v_{\rm rot}/\sigma$) for the same relaxed axisymmetric model (Fig.~\ref{vrot}, bottom) correspondent to the previous figures. This parameter shows the relative importance of rotational vs. pressure supported kinematics as used in observational studies of ellipticals and galaxy bulges of spirals. Here we used velocity bins of 50 stars and averaged over 50 time steps in order to get a better defined profile. In relaxed systems, the initial peak at $\sim r_{\rm hm}$ still dominates and becomes wider. Although the profile continuously decreases towards the center, an enhanced $v_{\rm rot}/\sigma$ inside $r_{\rm n}$ with respect to the initial configuration can be observed. It is specially interesting since the BH gravitational potential builds a velocity dispersion cusp ($\propto r^{-1/2}$), which requires a strong increase of $v_{\rm rot}$ in order to be detected.

\section{Conclusions}
Loss cone theory as developed in the classic papers of \cite{frank76, ls77} and \cite{cohn78} can be used to estimate feeding rates for SMBHs in galactic nuclei. Nonetheless, total consumption occurs from orbits that could extend beyond the black hole influence radius, hence the contribution of the stars to the gravitational potential cannot be ignored. These orbits interact with central stars and are able to interchange energy and angular momentum in relaxation time scales. In this work, we investigate accretion rates in axisymmetric systems by using direct NBody and Fokker-Planck simulations, harboring a star accreting growing black hole, which evolves dynamically in time scales of relaxation.

Our main results are:

\begin{itemize}

\item  The stellar distribution is strong influenced by the interplay between diffusion of energy and angular momentum (gravitational instabilities) and the feeding of the black hole influence zone by stars in high energy, low $J$ (eccentric) orbits. These systems undergo core collapse (reach a central density maximum at $t_{\rm cc}$) in the presence of a star-accreting black hole. The growing central BH potential dominates always larger zones of the system, with a growing influence radius, which reaches almost the half-mass radius during the post-collapse phase.

Axisymmetric (rotating) systems, like the spherically symmetric, reach in our simulations steady-state solutions during the post-collapse phase. Systems with smaller cores ($W_0=6.0$) reach self-similar expansion in shorter times than systems with larger cores ($W_0=3.0$). For relaxed systems, the BH $r_{\rm walk}$ is about one order of magnitude smaller than $r_{\rm crit}$, which is well resolved in our simulations, and itself about two orders of magnitude smaller than $r_{\rm n}$. Thus, these systems fulfill the conditions for the loss cone theory.

Furthermore, we are currently investigating the consequences for relaxation times and specially for the loss cone theory in multi-mass models. Mass segregation time scales in real systems are short enough to lead to a faster relaxation in the central parts, where the more massive stars concentrate, and consequently, to an earlier formation of cusps by these central stars (a work to be presented in a forthcoming publication). This is specially important, since galactic nuclei are often less than one relaxation time old, and stellar density near the SMBH seems not to have the Bahcall-Wolf form, although observations need to resolve the influence radius to be able to detect them. Additionally, relaxation times can be themselves much shorter, the smaller the radius of the system, and the higher the central densities.

\item We measured tidal disruption rates in axisymmetric NBody models and compare them to 2D Fokker-Planck realizations. Tidal star accretion acts as an indirect heating source reversing core collapse, like in isolated star clusters it is the role of hard binaries. In our low N, NBody runs with high initial seed black hole masses the indirect heating is stronger, so the density maximum, and thus the peak rates are wide. For high N or small seed black hole masses the heating is small, a very peaked central density and disruption rates occur (like in the FP models), as shown in Fig.~\ref{disrate}.

In order to investigate self-similar evolution during the post-collapse phase, we apply a scaling procedure to our models, which use partly different initial conditions and we obtain a non-trivial result for the rates through all the post-collapse phase where NBody and Fokker-Planck models agree with each other. This means that in this phase we can from these data make predictions for stellar disruption rates and other kinematical parameters of dry galactic nuclei, which are independent of the previous history (as also shown in \cite{fiestas10}), even independent of whether the black hole has grown earlier by gas or star accretion. A statement, which is confirmed here by using direct NBody realizations. 

We found that disruption rates and BH masses are influenced by axisymmetry/rotation, in the way that rotation leads to higher peak rates and higher $M_{\bullet,f}$. The excess of accreted stars, origins mainly from prograde rotating stars, located in regions outside $r_{\rm crit}$. This hints to a breakdown of the classical theory, given by the rotating models, which could be interpreted as a consequence of the presence of box orbits with non-conserved $J_{\rm x}$,$J_{\rm y}$ angular momentum.  To further investigate this effect, an orbit study of the accreted stars is necessary. This is of importance, since galactic nuclei need not be even axisymmetric. In a triaxial nucleus containing centrophilic orbits, the mass in stars on orbits that intersect the SMBH's capture sphere can be enormous, much greater than $M_\bullet$, so that the loss cone is never fully depleted. Galactic nuclei sometimes undergo catastrophic changes, due to galaxy mergers, in-fall of star clusters or black holes, star formation, etc, all of which can substantially affect the feeding rate on both the short and long terms.

We apply, for illustration, disruption rates given by Eq.~\ref{disr3} to the galactic center, by using $M_\bullet = 3.3 \times 10^6$, $r_{\rm n}= 1.65$, $\rho_0=2.8 \times 10^6 M_\odot {\rm pc}^{-3}$, $r_0=0.22 \ {\rm pc}$ and $\eta=1.75$ \citep{schoedel07}, and obtain stellar disruption rates of  $\sim 1.2 \times 10^{-4} M_\odot {\rm yr}^{-1}$. Our results presented in Table~\ref{results} give around 50 \% higher rates in axisymmetric models, which lead to final BH masses in average 20 \% higher with respect to spherically symmetric systems. This factor, would change the rates to $\sim 1.8 \times 10^{-4} M_\odot {\rm yr}^{-1}$. Integrated over the age of the universe, the mass gain is of the order of the BH mass. It means, that relaxation processes might play an important role in the growth of the galactic center BH. Moreover, for a galaxy core of $M \approx 10^9 M_\odot$ and a relaxation time of 10 Gyr, our models give us peak rates of the order of $\sim 10^{-4} M_\odot {\rm yr}^{-1}$. These rates are comparable to the accretion rates of some power law galaxies found by \cite{wang04}. 

\item Evolution of initially rotating systems with black holes affects substantially the orbital distribution of stars, specially in the regions inside the BH influence radius. We have found that rotation in relaxation time scales can not be neglected, and it triggers higher disruption rates with an excess of rotating stars. Central rotation has been detected in relaxation time scales in the zone of influence of the BH ($r_{\rm crit} < r < r_{\rm n}$), being at this time $r_{\rm n} \sim r_{\rm hm}(0)$. The original $v_{\rm rot,max}$ moves inwards, in a region where the BH potential dominates the stellar environment. For comparison, in systems without BH, thus without post-collapse evolution, dynamical instabilities cause that $v_{\rm rot,max}$ moves outwards from the center \citep{kim08}. In the central profile of the parameter $v_{\rm rot}/\sigma$ this maximum is lowered, mainly due to the cusp in the central velocity dispersion, but is not negligible.

The presence of different stellar populations is expected to enhance this effect in the central regions, as showed by \cite{kim04}. More massive stars segregate to the center in time scales shorter than a relaxation time, and they can rotate faster. Our currently investigations of multi-mass axisymmetric cores with stellar evolution and BHs, aim to obtain more detailed measurements of different stellar populations in the center, which could be detectable by observations. Another task, would be the treatment of a binary black hole (BBH), which can in a similar way, leads to a more efficient support in the development of rotation in its zone of influence \citep{berczik06,berentzen09}.
\end{itemize}

More realistic NBody simulations by using $\alpha=1$ and higher particle numbers up to $N \sim 10^{6}$ or more, are nevertheless necessary and still challenging to perform, specially in relaxation time scales, but will be possible in the near future. The advantage of using direct NBody models, together with computationally faster Fokker-Planck realizations makes possible to study the evolution of kinematical and structural parameters in more detail, which can complement and test observational measurements. Observational studies of 'collisional' galactic nuclei embedding massive black holes, can be compared to evolutionary models to elucidate theoretical predictions and have a better understanding of galaxy evolution.

\section*{Acknowledgments}
We acknowledge support by the Chinese Academy of Sciences Visiting Professorship for Senior International Scientists, Grant Number 2009S1-5 (The Silk Road Project) (RS,PB, and JF partly). The special supercomputer Laohu at the High Performance Computing Center at National Astronomical Observatories, funded by Ministry of Finance under the grant ZDYZ2008-2, has been used. Simulations were performed on the GRACE supercomputer (grants I/80 041-043 of the Volkswagen Foundation and 823.219-439/30 and /36 of the Ministry of Science, Research and the Arts of Baden-W\"urttemberg). 
The Kolob cluster is funded by the excellence funds of the University of Heidelberg in the Frontier scheme. We thank the {\bf SPP 1177} (SP 345/17-2) for the financial support of this project.
P.B. acknowledges the special support by the NAS Ukraine under the Main Astronomical Observatory GRAPE/GRID computing cluster project. P.B.'s studies are also partially supported by the program Cosmomicrophysics of NAS Ukraine. We thank the DEISA Consortium (http://www.deisa.eu), cofunded through EU FP6 projects RI-508830 and RI-031513, for support within the DEISA Extreme Computing Initiative. We thank the referee for fruitful comments, which helped to improve the quality of the present publication.


\begin{thebibliography}{99}
\bibitem[\protect\citeauthoryear{Aarseth}{1973}]{aarseth73} Aarseth S., 1973, VA, 15, 13
\bibitem[\protect\citeauthoryear{Ahmad \& Cohen}{1973}]{ahmad73} Ahmad A. \& Cohen L., JCoPh, 1973, 12, 389
\bibitem[\protect\citeauthoryear{Amaro-Seoane et al.}{2004}]{amaro04} Amaro-Seoane P., Freitag M. \& Spurzem R., 2004, \MN, 352, 655
\bibitem[\protect\citeauthoryear{Athanassoula, Vozikis \& Lambert}{2001}]{atha01} Athanassoula E., Vozikis C.~L., Lambert J.~C., 2001, \AA, 376, 1135 

\bibitem[\protect\citeauthoryear{Bahcall \& Wolf}{1977}]{bahcall77} Bahcall J. N. \& Wolf R. A., 1977, \ApJ, 216, 883
\bibitem[\protect\citeauthoryear{Baumgardt et al.}{2004}]{baumgardt04}  Baumgardt H., Portegies Zwart S. F., McMillan S. L. W. , Makino, J. \& Ebisuzaki T., 2004, ASPC, 322, 459
\bibitem[\protect\citeauthoryear{Baumgardt et al.}{2006}]{baumgardt06} Baumgardt H., Hopman C., Portegies Zwart S. \& Makino, J., 2006, \MN, 372, 467
\bibitem[\protect\citeauthoryear{Begelman et al.}{1980}]{begelman80} Begelman M.C., Blandford R.D. \& Rees M. J., 1980, \Nat, 287, 307                           
\bibitem[\protect\citeauthoryear{Berczik, Merritt, \& Spurzem}{2005}]{berczik05} Berczik P., Merritt D., Spurzem R., 2005, \ApJ, 633, 680 
\bibitem[\protect\citeauthoryear{Berczik et al.}{2006}]{berczik06} Berczik P., Merritt D., Spurzem R. \& Bischof H., 2006, \ApJ, 642, 21
\bibitem[\protect\citeauthoryear{Berczik et al.}{2011}]{berczik11} Berczik, P., Hamada, T., Spurzem, R., Berentzen, I., Preto, M., Merging of Unequal Mass binary black holes in non-axisymmetric galactic nuclei, to be submitted to \ApJ, 2011.
\bibitem[\protect\citeauthoryear{Berentzen et al.}{2009}]{berentzen09}  Berentzen I., Preto M., Berczik P., Merritt D., Spurzem R., 2009, \ApJ, 695,  455 
\bibitem[\protect\citeauthoryear{Callegari et al.}{2010}]{callegari10} Callegari S., Kazantzidis S., Mayer L.,Colpi M., Bellovary J.~M., Quinn T., Wadsley J., 2010, arXiv,arXiv:1002.1712
\bibitem[\protect\citeauthoryear{Callegari et al.}{2009}]{callegari09} Callegari S., Mayer L., Kazantzidis S.,Colpi M., Governato F., Quinn T., Wadsley J., 2009, \ApJ, 696, L89
\bibitem[\protect\citeauthoryear{Ciotti, Ostriker,\& Proga}{2010}]{ciotti10} Ciotti L., Ostriker J.~P., Proga D., 2010, \ApJ, 717, 708
\bibitem[\protect\citeauthoryear{Ciotti, Ostriker,\& Proga}{2009}]{ciotti09} Ciotti L., Ostriker J.~P., Proga D., 2009, \ApJ, 699, 89
\bibitem[\protect\citeauthoryear{Chatterjee, Hernquist \& Loeb}{2002}]{chatterjee02} Chatterjee P., Hernquist L. \& Loeb A., 2002, \PhRL, 88, 1103
\bibitem[\protect\citeauthoryear{Chen et al.}{2009}]{chen09} Chen X., Madau P., Sesana A., Liu F.~K., 2009, \ApJ, 697, L149 
\bibitem[\protect\citeauthoryear{Chen, Liu, \& Magorrian}{2008}]{chen08} Chen X., Liu F.~K., Magorrian J., 2008, \ApJ, 676, 54
\bibitem[\protect\citeauthoryear{Chen et al.}{2010}]{chen10} Chen X., Sesana A., Madau P., Liu F., 2010, arXiv, arXiv:1012.4466
\bibitem[\protect\citeauthoryear{Ciotti et al.}{2009}]{ciotti09} Ciotti L., Ostriker J.P. \& Proga D., 2009, \ApJ, 699, 89
\bibitem[\protect\citeauthoryear{Ciotti et al.}{2010}]{ciotti10} Ciotti L., Ostriker J.P. \& Proga D., 2010, arXiv, arXiv:1003.0578
\bibitem[\protect\citeauthoryear{Cohn \& Kulsrud}{1978}]{cohn78} Cohn H. \& Kulsrud R. M., 1978, \ApJ, 226, 1087
\bibitem[\protect\citeauthoryear{C{\^o}t{\'e} et al.}{2007}]{cote07} C{\^o}t{\'e} P., et al., 2007, \ApJ, 671, 1456 
\bibitem[\protect\citeauthoryear{Degraf et al.}{2010}]{degraf10}Degraf C, Di Matteo C. \& Springel V., 2010, \MN, 402, 1927
\bibitem[\protect\citeauthoryear{Dale et al.}{2009}]{dale09} Dale J.~E., Davies M.~B., Church R.~P., Freitag M., 2009, \MN, 393, 1016 
\bibitem[\protect\citeauthoryear{Do et al.}{2009}]{do09} Do T., Ghez A.~M., Morris M.~R., Yelda S., Meyer L., Lu J.~R., Hornstein S.~D., Matthews K., 2009, \ApJ, 691, 1021 
\bibitem[\protect\citeauthoryear{Dotti et al.}{2009}]{dotti09} Dotti M., Ruszkowski M., Paredi L., Colpi M., Volonteri M., Haardt F., 2009, \MN, 396, 1640 
\bibitem[\protect\citeauthoryear{Eilon, Kupi \& Alexander}{2009}]{eilon09} Eilon E., Kupi G., Alexander T., 2009, \ApJ, 698, 641 
\bibitem[\protect\citeauthoryear{Einsel \& Spurzem}{1999}]{einsel99} Einsel C. \& Spurzem R., 1999, \MN, 302, 81
\bibitem[\protect\citeauthoryear{Faber et al.}{1997}]{faber97} Faber S. M. et al., 1997, \AJ, 114, 1771
\bibitem[\protect\citeauthoryear{Ferrarese \& Ford}{2005}]{ferrarese05} Ferrarese L. \& Ford H., 2005, Space Science Reviews, Vol 116, Issue 3-4, pp. 523-624
\bibitem[\protect\citeauthoryear{Ferrarese et al.}{2006}]{ferrarese06} Ferrarese L., et al., 2006, \ApJ, 644, L21
\bibitem[\protect\citeauthoryear{Ferrarese et al.}{2006}]{ferrarese06} Ferrarese, L., et al.\ 2006, \ApJS, 164, 334 
\bibitem[\protect\citeauthoryear{Fiestas \& Spurzem}{2010}]{fiestas10} Fiestas J., Spurzem R., 2010, \MN, 405, 194 
\bibitem[\protect\citeauthoryear{Filippenko \& Ho}{2003}]{filippenko03} Filippenko A. \& Ho L. C., 2003, \ApJ, 588, 13
\bibitem[\protect\citeauthoryear{Frank \& Rees}{1976}]{frank76} Frank J. \& Rees M., 1976, \MN, 176, 633
\bibitem[\protect\citeauthoryear{Freitag \& Benz}{2002}]{freitag02}Freitag M. \& Benz W., 2002, \AA., 394, 345
\bibitem[\protect\citeauthoryear{Glass et al.}{2011}]{glass11} Glass L., et al., 2011, \ApJ, 726, 31 
\bibitem[\protect\citeauthoryear{Giersz \& Heggie}{1994}]{heggie94} Giersz M., Heggie D.~C., 1994, \MN, 268, 257 
\bibitem[\protect\citeauthoryear{Goodman}{1984}]{goodman84} Goodman J., 1984,  \ApJ, 280, 298
\bibitem[\protect\citeauthoryear{Gould \& Rix}{2000}]{gould00} Gould A., Rix H.-W., 2000, \ApJ, 532, L29 
\bibitem[\protect\citeauthoryear{Harfst et al.}{2007}]{harfst07} Harfst S., Gualandris A., Merritt D., Spurzem R., Portegies Zwart S., Berczik P., 2007, \NewA, 12, 357 
\bibitem[\protect\citeauthoryear{Heggie \& Mathieu}{1986}] {heggie86} Heggie,  D. C.; Mathieu, R. D., 1986, LNP, 267, 233
\bibitem[\protect\citeauthoryear{Hemsendorf, Sigurdsson, \& Spurzem}{2002}]{hemsendorf02} Hemsendorf M., Sigurdsson S., Spurzem R., 2002, \ApJ, 581, 1256 
\bibitem[\protect\citeauthoryear{H\'{e}non}{1965}] {henon65} H\'{e}non M., 1965, AnAp, 28, 62
\bibitem[\protect\citeauthoryear{Hirschmann et al.}{2010}]{hirschmann10} Hirschmann M., Khochfar S., Burkert A.,Naab T., Genel S., Somerville R.~S., 2010, \MN, 407, 1016
\bibitem[\protect\citeauthoryear{Hopman \& Alexander}{2006}] {hopman06} Hopman C. \& Alexander T., 2006, \ApJ, 645, 1152
\bibitem[\protect\citeauthoryear{Hurley et al}{2000}] {hurley00} Hurley J. R., Pols O. R. \& Tout C. A., 2000, \MN, 315, 543
\bibitem[\protect\citeauthoryear{Johansson, Burkert,\& Naab}{2009}]{johansson09a} Johansson P.~H., Burkert A., Naab T., 2009, \ApJ, 707, L184
\bibitem[\protect\citeauthoryear{Johansson, Naab,\& Burkert}{2009}]{johansson09b} Johansson P.~H., Naab T., Burkert A., 2009, \ApJ, 690, 802
\bibitem[\protect\citeauthoryear{Just et al.}{2011}]{just11} Just A., Yurin  D., Makukov M., Berczik P., Omarov C., Spurzem R., Vilkoviski E.Y., Enhanced  accretion rates on Super-massive Black Holes by star-disk interactions in  galactic nuclei, subm. to \ApJ 2011.
\bibitem[\protect\citeauthoryear{Kim, Lee, \& Spurzem}{2004}]{kim04} Kim E., Lee H.~M., Spurzem R., 2004, \MN, 351, 220 
\bibitem[\protect\citeauthoryear{Kim et al.}{2008}] {kim08} Kim E., Yoon I., Lee H.M. \& Spurzem R., 2008, \MN, 383, 2
\bibitem[\protect\citeauthoryear{King}{1962}] {king62} King I., 1962, \AJ, 67, 471
\bibitem[\protect\citeauthoryear{Kobayashi}{2004}] {kobayashi04} Kobayashi S., Laguna P., Phinney E. S. \& M市zõ€'šos P., 2004, \ApJ, 615, 855
\bibitem[\protect\citeauthoryear{Khokhlov \& Melia}{1996}] {khokhlov96} Khokhlov A. \& Melia F., 1996, \ApJ, 457, 61
\bibitem[\protect\citeauthoryear{Komossa et al.}{2004}] {komossa04} Komossa  S., Halpern J., Schartel N., Hasinger G., Santos-Lleo M. \& Predehl P.,  2004, \ApJ, 603, 17
\bibitem[\protect\citeauthoryear{Kormendy \& Richstone}{1995}] {kormendy95} Kormendy J. \& Richstone D., 1995, \ARAA, 33, 581
\bibitem[\protect\citeauthoryear{Lauer et al.}{1998}]{lauer98} Lauer T.~R., Faber S. M., Ajhar E. A., Grillmair C. J. \& Scowen P. A., 1998, \AJ, 116, 2263
\bibitem[\protect\citeauthoryear{Lightman \& Shapiro}{1977}]{ls77} Lightman, A.ツ P. \& Shapiro, S.ツ L. 1977, \ApJ, 211, 244 
\bibitem[\protect\citeauthoryear{Liu, Li, \& Chen}{2009}]{liu09} Liu F.~K., Li S., Chen X., 2009, \ApJ, 706, L133
\bibitem[\protect\citeauthoryear{Madau \& Rees}{2001}]{MR-01} Madau P. \& Rees M.~J., 2001, \ApJ, 551, L27 
\bibitem[\protect\citeauthoryear{Madigan, et al.}{2011}]{madigan11} Madigan A.-M., Hopman C., Levin Y., 2011, \ApJ, 738,99 
\bibitem[\protect\citeauthoryear{Magorrian \& Tremaine}{1999}]{magorrian99} Magorrian J. \& Tremaine S., 1999, \MN, 309, 447
\bibitem[\protect\citeauthoryear{Makino \& Funato}{2004}]{makino04} Makino J., Funato Y., 2004, \ApJ, 602, 93 
\bibitem[\protect\citeauthoryear{Makino \& Sugimoto}{1987}]{makino87} Makino J., Sugimoto D., 1987, \PASJ, 39, 589 
\bibitem[\protect\citeauthoryear{Mapelli et al.}{2010}]{mapelli10} Mapelli M.,  Huwyler C., Mayer L., Jetzer P., Vecchio A., 2010, \ApJ, 719, 987 
\bibitem[\protect\citeauthoryear{Mayer et al.}{2010}]{mayer10} Mayer L., Kazantzidis S., Escala A., Callegari S., 2010, \Nat, 466, 1082
\bibitem[\protect\citeauthoryear{Marchant \& Shapiro}{1980}]{marchant80} Marchant A. B. \& Shapiro S. L., 1980, \ApJ, 239, 685
\bibitem[\protect\citeauthoryear{Merritt, Berczik \& Laun}{2007}]{merritt07} Merritt D., Berczik P., Laun F., 2007, \AJ, 133, 553 
\bibitem[\protect\citeauthoryear{Merritt \& Poon}{2004}]{merritt04} Merritt D. \& Poon M. Y., 2004, \ApJ, 606, 788
\bibitem[\protect\citeauthoryear{Merritt \& Szell}{2006}]{merritt06} Merritt D. \& Szell A., 2006, \ApJ, 648, 890 
\bibitem[\protect\citeauthoryear{Merritt}{2010}]{merritt10} Merritt D., 2010, \ApJ, 718, 739 
\bibitem[\protect\citeauthoryear{Merritt }{2006}]{RPP06} Merritt D., 2006, Rep. Prog. Phys., 69, 2513
\bibitem[\protect\citeauthoryear{Milosavljevi{\'c} \& Merritt}{2003}]{milos03} Milosavljevi{\'c} M., Merritt D., 2003, ApJ, 596, 860 
\bibitem[\protect\citeauthoryear{Milosavljevi\'{c} et al.}{2006}]{milos06} Milosavljevi\'{c} M., Merritt D. \& Ho L., 2006, \ApJ,  652, 120
\bibitem[\protect\citeauthoryear{Murphy et al.}{1991}]{murphy91} Murphy B.W., Cohn H.N. \& Durisen R.H., 1991, \ApJ, 370, 60
\bibitem[\protect\citeauthoryear{Norman \& Silk}{1983}]{norman83} Norman C., Silk J., 1983, \ApJ, 266, 502 
\bibitem[\protect\citeauthoryear{Peebles}{1972}] {peebles72} Peebles P.J.E., 1972, \ApJ, 178, 37
\bibitem[\protect\citeauthoryear{Perets \& Alexander}{2008}]{perets08} Perets H.~B., Alexander T., 2008, \ApJ, 677, 146 
\bibitem[\protect\citeauthoryear{Portegies Zwart et al.}{2004}]{portegies04} Portegies Zwart S.F., Baumgardt H., Hut P., Makino J. \& McMillan S.L., 2004, \Nat, 428, 724
\bibitem[\protect\citeauthoryear{Preto, Merritt \& Spurzem}{2004}]{preto04} Preto M., Merritt D. \& Spurzem R., \ApJ, 2004, 613, 109
\bibitem[\protect\citeauthoryear{Rees}{1988}]{rees88} Rees M.~J., 1988, \Nat, 333, 523 
\bibitem[\protect\citeauthoryear{Rodriguez et al.}{2006}]{rodriguez06} Rodriguez C., Taylor G.B., Zavala R.T., Peck A.B., Pollack L.K. \& Romani R. W., 2006, \ApJ, 646, 49
\bibitem[\protect\citeauthoryear{Scannapieco et al.}{2005}]{scannapieco05} Scannapieco E., Silk J. \& Bouwens R., 2005, \ApJ, 635, 13
\bibitem[\protect\citeauthoryear{Sch{\"o}del et al.}{2007}]{schoedel07} Sch{\"o}del R., et al., 2007, \AA, 469, 125 
\bibitem[\protect\citeauthoryear{Sesana}{2010}]{sesana10} Sesana A., 2010, \ApJ, 719, 851
\bibitem[\protect\citeauthoryear{Shankar}{2009}]{shankar09} Shankar F., 2009, New Astronomy Review, 53, 57
\bibitem[\protect\citeauthoryear{Shapiro}{1977}]{shapiro77} Shapiro S.~L.,  1977, \ApJ, 217, 281
\bibitem[\protect\citeauthoryear{Shin, Ostriker \& Ciotti}{2010}]{shin10a} Shin M.-S., Ostriker J.~P., Ciotti L., 2010, arXiv, arXiv:1003.1108
\bibitem[\protect\citeauthoryear{Shin, Ostriker,\& Ciotti}{2010}]{shin10b} Shin M.-S., Ostriker J.~P., Ciotti L., 2010, \ApJ, 711, 268
\bibitem[\protect\citeauthoryear{Shin et al.}{2010}]{shin10} Shin M-S., Ostriker J.P. \& Ciotti L., 2010, arXiv, arXiv:1003.1108 
\bibitem[\protect\citeauthoryear{Spitzer}{1987}]{spitzer87}, Spitzer, 1987, degc.book.....S
\bibitem[\protect\citeauthoryear{Spurzem}{1999}]{spurzem99} Spurzem R., 1999, JCoAM, 109, 407
\bibitem[\protect\citeauthoryear{Toomre}{1977}]{toomre77} Toomre A., 1977, in ``Evolution of Galaxies and Stellar Populations,'' ed. B. M. Tinsley and R. B. Larson. (New Haven: Yale University Observatory), p.401
 \bibitem[\protect\citeauthoryear{Valluri et al.}{2005}]{valluri05} Valluri M., Ferrarese L., Merritt D. \& Joseph Ch.L., 2005, \ApJ, 628, 137
\bibitem[\protect\citeauthoryear{Valtonen et al.}{2008}]{valtonen08} Valtonen M.~J., et al., 2008, \Nat, 452, 851 
\bibitem[\protect\citeauthoryear{van der Marel}{2004}]{vandermarel04} van der Marel R. P., 2004, cbhg.symp, 37
\bibitem[\protect\citeauthoryear{Vilkoviskij \& Czerny}{2002}]{vilkoviski02} Vilkoviskij E.~Y., Czerny B., 2002, \AA, 387, 804 
\bibitem[\protect\citeauthoryear{Wang \& Merritt}{2004}] {wang04} Wang J. \& Merritt D., 2004, \ApJ, 600, 149
\bibitem[\protect\citeauthoryear{Willot et al.}{2010}] {willot10} Willot C.J., Delorme P., Rey....C., Albert L., Bergeron J., Crampton D., Delfosse X., Forveille T., Hutchings J.B., McLure R.J., Omont A. \& Schade D., 2010, \AJ, 139, 906
\bibitem[\protect\citeauthoryear{Yu \& Tremaine}{2002}]{yu02} Yu Q., Tremaine S., 2002,\MN, 335, 965 

\end{thebibliography}
\end{document}